\documentclass{aa}
\usepackage{natbib}
\usepackage{pdflscape} 
\usepackage{wasysym}
\usepackage[varg]{txfonts}
\usepackage{graphicx}
\usepackage[draft]{hyperref}
\bibpunct{(}{)}{;}{a}{}{,} 
\usepackage{longtable}
\usepackage{amssymb}

\usepackage{afterpage}

\def\ms{\hbox{\,m\,s$^{-1}$}}         
\def\m2s2{\hbox{\,m$^{2}$\,s$^{-2}$}} 
\def\Msun{\hbox{$\mathrm{M}_{\odot}$}}             
\def\Rsun{\hbox{$\mathrm{R}_{\odot}$}}
\def\Mearth{\hbox{$\mathrm{M}_{\oplus}$}}             
\def\Rearth{\hbox{$\mathrm{R}_{\oplus}$}}

\def\Rjup{\hbox{$\mathrm{R}_{\rm Jup}$}}

\def\logrhk{$\log(R^{\prime}_\mathrm{HK}$)}

\newcommand{\Zenv}{$Z_{\rm gas}$\xspace}
\newcommand{\Lenv}{$L_{\rm int}$\xspace}

\newcommand{\rsolid}{$r_{\rm core+mantle}$\xspace}

\newcommand{\fesima}{{\rm Fe}/{\rm Si}_{\rm mantle}\xspace}
\newcommand{\mgsima}{{\rm Mg}/{\rm Si}_{\rm mantle}\xspace}

\usepackage{xspace}

\usepackage{xcolor}

\begin{document}

\title{Hot, rocky and warm, puffy super-Earths orbiting TOI-402 (HD\,15337)
\thanks{Based on observations made with the HARPS instrument on the ESO 3.6 m telescope at La Silla Observatory under the GTO program 072.C-0488(E) and Large programs 183.C-0972(A), 192.C-0852(A), 196.C-1006 and 198.C-0836(A).}
\thanks{The HARPS RV measurements discussed in this paper are available in electronic form at the CDS via anonymous ftp to cdsarc.u-strasbg.fr (130.79.128.5) or via http://cdsweb.u-strasbg.fr/cgi-bin/qcat?J/A+A/.}
}
\author{Xavier~Dumusque\inst{1}
            \and Oliver~Turner\inst{1}
            \and Caroline Dorn\inst{2}
            \and Jason D.~Eastman\inst{3}
            \and Romain~Allart\inst{1}
            \and Vardan~Adibekyan\inst{4}
            \and Sergio~Sousa\inst{4}
            \and Nuno C.~Santos\inst{4,5}
            \and Christoph Mordasini\inst{6}
            \and Vincent Bourrier\inst{1}
            \and Fran\c cois~Bouchy\inst{1}
            \and Adrien~Coffinet\inst{1}
            \and Misty D.~Davies\inst{7}
            \and Rodrigo F.~D\'iaz\inst{8,9}
            \and Michael M.~Fausnaugh\inst{10}
            \and Ana~Glidden\inst{10,14}
            \and Natalia~Guerrero\inst{10}
            \and Christopher E.~Henze\inst{7}
            \and Jon M.~Jenkins\inst{7}
            \and David W.~Latham\inst{3}
            \and Cristophe~Lovis\inst{1}
            \and Michel~Mayor\inst{1}
            \and Francesco~Pepe\inst{1} 
            \and Elisa V.~Quintana\inst{11}
            \and George R.~Ricker\inst{10}
            \and Pamela~Rowden\inst{12}
            \and Damien~Segransan\inst{1}
            \and Alejandro~Su\'arez Mascare\~no\inst{13}
            \and Sara~Seager\inst{14}
            \and Joseph D.~Twicken\inst{7,15}
            \and St\'ephane~Udry\inst{1}
            \and Roland K.~Vanderspek\inst{10}
            \and Joshua N.~Winn\inst{16}
}
\offprints{Xavier Dumusque, \email{xavier.dumusque@unige.ch}}
\institute{Astronomy Department of the University of Geneva, 
               51 ch. des Maillettes, CH-1290 Versoix, Switzerland
}

\institute{
Astronomy Department of the University of Geneva, 51 chemin des Maillettes, CH-1290 Versoix, Switzerland.
\and 
University of Zurich, Institut of Computational Sciences, Winterthurerstrasse 190, CH-8057, Zurich, Switzerland.
\and
Harvard-Smithsonian Center for Astrophysics, 60 Garden St, Cambridge, MA 02138, USA
\and
Instituto de Astrof\'isica e Ci\^encias do Espa\c co, Universidade do Porto, CAUP, Rua das Estrelas, 4150-762 Porto, Portugal.
\and 
Departamento de F\'isica e Astronomia, Faculdade de Ci\^encias, Universidade do Porto, Rua do Campo Alegre, 4169-007 Porto, Portugal.
\and
Physikalisches Institut, Universitat Bern, Silderstrasse 5, 3012 Bern, Switzerland.
\and
NASA Ames Research Center, Moffett Field, CA, 94035, USA.
\and
Universidad de Buenos Aires, Facultad de Ciencias Exactas y Naturales. Buenos Aires, Argentina.
\and
CONICET - Universidad de Buenos Aires. Instituto de Astronom\'ia y F\'isica del Espacio (IAFE). Buenos Aires, Argentina.
\and
Department of Physics and Kavli Institute for Astrophysics and Space Research, Massachusetts Institute of Technology, Cambridge, MA 02139, USA.
\and
NASA Goddard Space Flight Center, 8800 Greenbelt Road, Greenbelt MD 20771.
\and
School of Physical Sciences, The Open University, Walton Hall, Milton Keynes MK7 6AA, United Kingdom.
\and
Instituto de Astrof\'isica de Canarias, 38205 La Laguna, Tenerife, Spain.
\and
Department of Earth, Atmospheric, and Planetary Sciences, Massachusetts Institute of Technology, Cambridge, MA 02139, USA
\and
SETI Institute, Mountain View, CA 94043, USA.
\and
Department of Astrophysical Sciences, 4 Ivy Lane, Princeton University, Princeton, NJ USA 08544.
}

\date{Received XXX; accepted XXX}

\abstract
{The Transiting Exoplanet Survey Satellite (\emph{TESS}) is revolutionising the search for planets orbiting bright and nearby stars. In sectors 3 and 4, \emph{TESS} observed TOI-402 (TIC-120896927), a bright V=9.1 K1 dwarf also known as \object{HD\,15337},
and found two transiting signals with periods of 4.76 and 17.18 days and radii of 1.90 and 2.21\,\Rearth, respectively. This star was observed prior to the TESS detection as part of the radial-velocity (RV) search for planets using the HARPS spectrometer, and 85 precise RV measurements were obtained before the launch of \emph{TESS} over a period of 14 years.}
{In this paper, we analyse the HARPS RV measurements in hand to confirm the planetary nature of these two signals.}
{\object{HD\,15337} happens to present a stellar activity level similar to the Sun, with a magnetic cycle of similar amplitude and RV measurements that are affected by stellar activity.
By modelling this stellar activity in the HARPS radial velocities using a linear dependence with the calcium activity index \logrhk, we are able, with a periodogram approach, to confirm the periods and 
the planetary nature of TOI-402.01 and TOI-402.02.
We then derive robust estimates  from the HARPS RVs for the orbital parameters of these two planets by modelling stellar activity with a Gaussian process and using the marginalised posterior probability density functions obtained from our analysis of \emph{TESS} photometry for the orbital period and time of transit.}
{By modelling TESS photometry and the stellar host characteristics, we find that TOI-402.01 and TOI-402.02 have periods of 4.75642$\pm$0.00021 and 17.1784$\pm$0.0016 days and radii of 
1.70$\pm$0.06 and 2.52$\pm$0.11\,\Rearth\,(precision 3.6 and 4.2\%), respectively. By analysing the HARPS RV measurements, we find that those planets are both super-Earths with masses of 7.20$\pm$0.81 and 
8.79$\pm$1.68\,\Mearth\,(precision 11.3 and 19.1\%), and small eccentricities compatible with zero at 2$\sigma$.
}
{Although having rather similar masses, the radii of these two planets are very different, putting them on different sides of the radius gap. 
By studying the temporal evolution under X-ray and UV (XUV) driven atmospheric escape of the TOI-402 planetary system, we confirm, under the given assumptions, that photo-evaporation is a plausible explanation
for this radius difference. Those two planets, being in the same system and therefore being
in the same irradiation environment are therefore extremely useful for comparative exoplanetology across the evaporation valley and thus bring constraints on the mechanisms responsible for the radius gap.}
   

\authorrunning{X. Dumusque}
\titlerunning{A hot rocky and a warm puffy Super-Earth orbiting TOI-402}

\maketitle

\section{Introduction}
\label{sec:intro}

The Transiting Exoplanet Survey Satellite \citep[\emph{TESS},][]{Ricker:2015aa} is revolutionising the field of exoplanets by detecting such bodies transiting bright and nearby stars. 
This allows for ground-based spectroscopic follow-up studies to be performed to first confirm those planets and then obtain their densities, which is needed to understand planetary formation. Planets transiting bright stars are also extremely interesting for ground- and space-based follow-up observations aimed at studying their atmospheres by using transmission or emission spectroscopy.

\emph{TESS} will observe nearly the whole sky in a two-year period, implying that most of the targets observed until now in blind radial-velocity (RV) surveys for exoplanets will be measured \citep[e.g. $\Pi$\,Men,][]{Huang:2018ab, Gandolfi:2018aa}.
This is the case for TOI-402 (TIC ID 120896927), also known as HD\,15337, for which a significant number of HARPS observations were obtained. \emph{TESS} found two transiting candidates
orbiting around this star, and the goal of this paper is to confirm them using the publicly available HARPS observations. Note that an independent analysis of the same planetary system using 
the same data can be found in \citep[][]{Gandolfi:2019aa}. The results from this paper are consistent with the analysis performed here.

In Sect.\,\ref{sec:obs}, we describe the photometric and spectroscopic observations obtained by \emph{TESS} and HARPS. In Sect.\,\ref{sec:analysis_results}, we analyse those data
to confirm the two planets found by \emph{TESS} and to obtain good constraints on their radius and mass. We finally discuss our results and conclude in Sect.\,\ref{sec:discussion}.

\section{Observations}
\label{sec:obs}

\subsection{TESS photometry}
\label{subsec:phot}

TOI-402 was observed by \emph{TESS} in two-minute cadence mode during sectors 3 and 4, that is between September 22 and November 15, 2018.
The target appeared on camera 2 and CCDs 3 and 4 for sectors 3 and 4, respectively.  Nine transits of TOI-402.01 and three of TOI-402.02 were detected in the two sectors, 
with depths of 406 and 609 ppm with a S/N of 10.4 and 10.7, respectively\footnote{We note that those S/N and transit depths are given by the Science Processing Operations Center (SPOC) pipeline transit search detection statistics and are based on sector 4 data only.} \citep[][]{Jenkins:2017ab, Twicken:2018aa}. 

In Sect.\,\ref{sec:analysis_TESS}, we analyse
the data publicly available on MAST (\url{https://mast.stsci.edu/portal/Mashup/Clients/Mast/Portal.html}), which were reduced by the 
Science Processing Operations Center pipeline \citep[][]{Jenkins:2016aa, Jenkins:2017aa, Jenkins:2017ab}.
For our transit modelling in Sect.\,\ref{sec:analysis_TESS}, we used the raw light curve based on the Pre-search Data Conditioning 
Simple Aperture Photometry \citep[PDCSAP,][]{Twicken:2010aa, Smith:2012aa, Stumpe:2014aa}. 

\subsection{High-resolution spectroscopy with HARPS}
\label{sec:rv_harps}

The star HD\,15337 was observed with HARPS on the ESO 3.6m telescope in La Silla, Chile, between December 2003 and September 2017.
Those observations were gathered in the framework of the HARPS guaranteed time observation (GTO) and subsequent high-precision large programs (LP) obtained on open time. We note that the GTO and subsequent LP observations lead to the discovery of a large number of super-Earths and Neptunes orbiting nearby solar-type stars \citep[e.g.][]{Lovis:2009aa, Mayor-2011, Udry:2019aa}.

Out of a total of 87 spectra available on the ESO archive, 85  passed the quality control of the HARPS Data Reduction Software (DRS). The S/N of the 
spectra at 550\,nm ranges between 47 and 142 for a photon-noise RV precision of 0.56 and 1.59\,\ms, respectively.  
The observations were performed with two different configurations of the HARPS instrument, with circular fibre before June 1, 2015 (BJD$_{UTC}=2457174.5$), and octagonal fibres after that. This change of optical fibres induced an offset in the RV time-series that depends on the stellar spectral type and is difficult to determine beforehand at the level of meter per second or below \citep[][]{Lo-Curto:2015aa} and therefore, it must be taken into account by adjusting an additional free parameter. In the RV data of HD\,15337, the difference between the median of the data before and after June 1, 2015, is 42.4\,\ms. After correcting for this offset, the RV residuals  are left with a RV rms of 4.37\,\ms, much higher than the photon noise uncertainty, pointing towards the presence of extra signals in the data of stellar or planetary origin.

We re-reduced the 85 spectra using the latest and very recent version of the DRS that includes a better wavelength solution owing to the consideration of the CCD stitching \citep[][]{Dumusque-2015a, Coffinet:2019aa} and publish the new extracted RVs and \logrhk\,at the CDS. \citet{Coffinet:2019aa} demonstrated that HARPS data derived with this new DRS are of much better quality. In the case of HD\,15337, this new version of the DRS gives very similar results to those from the previous one for the RV data before the change of the fibres (rms of the RV difference 0.3\,\ms); however, the differences are significant for the data gathered after, the rms of the RV difference being 1.6\,\ms. As the data after June 1, 2015, represent 40\% of the measurements, using the data derived with this latest version of the DRS has a significant impact on precision.

We note that all the HARPS publicly available data are currently being re-reduced with this latest version of the DRS at the Astronomy Department of the University of Geneva. Those data will be made available to the community through the Data and Analysis Center for Exoplanets web platform (DACE, \url{https://dace.unige.ch}).

\section{Analysis and Results}
 \label{sec:analysis_results}

\subsection{Stellar parameters and chemical abundances}           
 \label{sec:stellar_para}

The stellar atmospheric parameters ($T_{\rm eff}$, $\log{g}$, micro-turbulence and [Fe/H]) and respective error bars were derived using the methodology described in \citet{Sousa-14} and \citet{Santos-13}. We make use of the equivalent widths (EWs) of well-selected iron lines. For this relatively cool star we used the appropriate line list presented in \citet[][]{Tsantaki-13}. The equivalent widths were measured for the combined HARPS spectrum of TOI-402 using the ARES v2 code\footnote{The last version of ARES code (ARES v2) can be downloaded at http://www.astro.up.pt/$\sim$sousasag/ares} \citep{Sousa-07, Sousa-15}, and we assume ionisation and excitation equilibrium. The process makes use of a grid of Kurucz model atmospheres \citep{Kurucz-93} and the radiative-transfer code MOOG \citep{Sneden-73}. As discussed in the references above, this method provides effective temperatures in excellent agreement with values derived using the infra-red flux method \citep[][]{Casagrande:2006aa, Casagrande:2010aa} and which are independent of the derived surface gravity \citep[][]{Torres:2012aa}. The resulting values for $T_{\rm eff}$, $\log{g}$, micro-turbulence, and [Fe/H] can be found in Table\,\ref{tab:0}.

Stellar abundances of the elements were also derived using the same tools and models as for stellar parameter determination, as well as using the classical curve-of-growth analysis method assuming local thermodynamic equilibrium. Although the EWs of the spectral lines were automatically measured with ARES, for the elements with fewer than four lines available we performed careful visual inspection of the EW measurements. For the derivation of chemical abundances of refractory elements we closely followed the methods described in \citet[][]{ Adibekyan-15}. The final abundances derived, which are fully compatible with the ones found in the Hypatia catalogue \citep[\url{https://www.hypatiacatalog.com},][]{Hinkel:2014aa}, can be found in Table\,\ref{tab:0}. When comparing the chemical composition with stars of similar atmospheric parameters in the solar neighbourhood \citep[][]{Adibekyan-12}, we find that TOI-402 has typical properties for a thin disc star.
%
\renewcommand{\arraystretch}{1.3}
\begin{table}[ht]
        \scriptsize
        \caption{Stellar properties for TOI-402}            
        \label{tab:0}     
        \centering                         
        \begin{tabular}{ccc}       
                \hline\hline
                Property & Value & Reference\\
                \hline
                \multicolumn{3}{c}{\bf{Names}}\\        
                TOI      & TOI-402     & \emph{TESS}\\
                TIC ID & 120896927 & \emph{TESS}\\
                HD      & 15337 & --\\
                2MASS ID & J02272838-2738064 & 2MASS\\
                \emph{Gaia} ID DR2 & 5068777809824976256 & \emph{Gaia}\\
                \multicolumn{3}{c}{\bf{Astrometric properties}}\\
                Parallax (mas) & 22.2918$\pm$0.0349 & \emph{Gaia}\tablefootmark{+}\\ 
                Distance (pc) & 44.86$\pm$0.07 & \emph{Gaia}\tablefootmark{+}\\ 
                
                \multicolumn{3}{c}{\bf{Photometric properties}}\\
                B-V & 0.86 & Tycho\\
                B (mag) & 9.95$\pm$0.03 & Tycho\\
                V (mag) & 9.09$\pm$0.02 & Tycho\\
                G (mag) & 8.8560$\pm$0.0002 & \emph{Gaia}\\
                T (mag) & 8.27$\pm$0.02& \emph{TESS}\\
                J (mag) & 7.55$\pm$0.02& 2MASS\\
                H (mag) & 7.22$\pm$0.04& 2MASS\\
                K$_s$ (mag) & 7.04$\pm$0.02& 2MASS\\
                W1 (mag) & 6.92$\pm$0.05& WISE\\
                W2 (mag) & 7.05$\pm$0.02& WISE\\
                W3 (mag) & 7.01$\pm$0.02& WISE\\
                W4 (mag) & 6.92$\pm$0.07& WISE\\
                \multicolumn{3}{c}{\bf{Spectroscopic properties}}\\
                Spectral type  & K1V & \citet{Houk:1982aa}\\
                $T_{\rm eff}$ ($K$) & 5131$\pm$74 & see Sect.\,\ref{sec:stellar_para}\\
                $\log{g}$ & 4.37$\pm$0.13 & see Sect.\,\ref{sec:stellar_para}\\
                $\xi_{t}$ & 0.87$\pm$0.13 & see Sect.\,\ref{sec:stellar_para}\\
                $[Fe/H]$ & 0.03$\pm$0.04 & see Sect.\,\ref{sec:stellar_para}\\
                 $[NaI/H]$ & 0.12$\pm$0.08 & see Sect.\,\ref{sec:stellar_para}\\
                 $[MgI/H]$ & 0.10$\pm$0.07 & see Sect.\,\ref{sec:stellar_para}\\
                 $[AlI/H]$ & 0.11$\pm$0.06 & see Sect.\,\ref{sec:stellar_para}\\
                 $[SiI/H]$ & 0.05$\pm$0.07 & see Sect.\,\ref{sec:stellar_para}\\
                 $[CaI/H]$ & -0.03$\pm$0.10 & see Sect.\,\ref{sec:stellar_para}\\
                 $[ScII/H]$ & 0.04$\pm$0.08 & see Sect.\,\ref{sec:stellar_para}\\
                 $[TiI/H]$ & 0.10$\pm$0.10 & see Sect.\,\ref{sec:stellar_para}\\
                 $[CrI/H]$ & 0.05$\pm$0.08 & see Sect.\,\ref{sec:stellar_para}\\
                 $[NiI/H]$ & 0.07$\pm$0.04 & see Sect.\,\ref{sec:stellar_para}\\
                 \multicolumn{3}{c}{\bf{Bulk properties}}\\
                 Mass (\Msun) & 0.851$^{+0.042}_{-0.034}$ (4.9\%) & see Sect.\,\ref{sec:analysis_TESS}\\
                 Radius (\Rsun) & 0.839$^{+0.018}_{-0.016}$ & see Sect.\,\ref{sec:analysis_TESS}\\
                 Luminosity (L$_{\odot}$) & 0.472$^{+0.023}_{-0.024}$ & see Sect.\,\ref{sec:analysis_TESS}\\
                 Age (Gyr) & 7.5$^{+4.2}_{-4.4}$ & see Sect.\,\ref{sec:analysis_TESS}\\
                 \multicolumn{3}{c}{\bf{Activity}}\\
                 $<$\logrhk$>$ & -4.91 & see Sect.\,\ref{sec:stellar_rot}\\
                 P$_{rot}$ (days) & 36.55 & see Sect.\,\ref{sec:stellar_rot}\\
                \hline
        \end{tabular}
        \tablefoot{
        \tablefoottext{+} For completeness, we adjusted the parallax by +0.07 mas based on the systematic offsets reported by \citet{Stassun:2018aa} and by \citet{Zinn:2018aa}. Both studies find the Gaia DR2 parallaxes to be too small by 0.06$\sim$0.08 mas from benchmark eclipsing binary and asteroseismic stars within $\sim$1 kpc.
        }
\end{table}
%

\subsection{Analysis of \emph{TESS} photometry}           
 \label{sec:analysis_TESS}

We performed the analysis of the \emph{TESS} photometry using EXOFASTv2 \citep{Eastman:2017ascl,Eastman:2013PASP} 
to jointly model the light curve and the host star. We imposed priors on the stellar model generated by Mesa isochrones and stellar tracks \citep[MIST,][] {Dotter:2016ApJS,Choi:2016ApJ} using the spectral parameters derived in Sect. \ref{sec:stellar_para}, the photometry in the spectral band-passes detailed in Table\,\ref{tab:0}, and the Gaia parallax shown in the same table. 

We removed the long-timescale variations due to stellar activity from the \emph{TESS} light curve of TOI-402 using a moving polynomial, as applied for K2 photometry by \citet{Giles:2018MNRAS}. This method consists of fitting a sliding polynomial which at each sliding step fits a large fraction of the light curve ("window") but only divides this model out of a small section ("step size"). This prevents the method from being sensitive to side effects common with polynomials. In our analysis, we used a third-order polynomial with a step size of 0.1 day and a window size of 5 days. To ensure the result is not jagged, the step size must be significantly smaller than the window size. To preserve the transit depth during this smoothing process, and therefore prevent biasing the final derived planetary radius, we need to perform outlier rejection from the polynomial fit. This is done by doing a 3$\sigma$ cut of positive outliers and selecting a X$\sigma$ cut on negative outliers, where X is chosen so that X$\sigma$ is larger than the deepest transit. 


When fitting a light curve, EXOFASTv2 uses 31 free parameters. We placed Gaussian priors on $T_{\rm eff}$, log($g$), [Fe/H], and parallax based on our spectral analysis and \emph{Gaia} results (see Sect.\,\ref{sec:stellar_para} and Table\,\ref{tab:0}). We placed loose constraints on the period, epoch, and transit depth to prevent the Monte Carlo Markov Chain (MCMC) sampler from entering poor fitting areas of parameter space and left the remaining parameters free. The planet orbits are considered non-eccentric. 

For the final MCMC analysis run we used the output values from an initial analysis as starting values. We used a maximum of 50,000 steps and up to 50 independent chains. While running, EXOFASTv2 uses the Gelman-Rubin (GR) statistic \citep[][]{Gelman:1992aa}, which compares the within and between chain variances to assess convergence, and $T_{\rm z}$, the number of independent draws to assess if the chains have converged. Upon reaching a GR $<1.01$ and $T_{\rm z}>1000$ for a sufficient number of draws, EXOFASTv2 considers the chains converged and stops. In our analysis, this was reached after 27,500 steps. To get the final marginalised posterior probability function densities for each parameter, the chains were merged together.

\begin{figure*}[h]
        \centering
        \includegraphics[width=6cm]{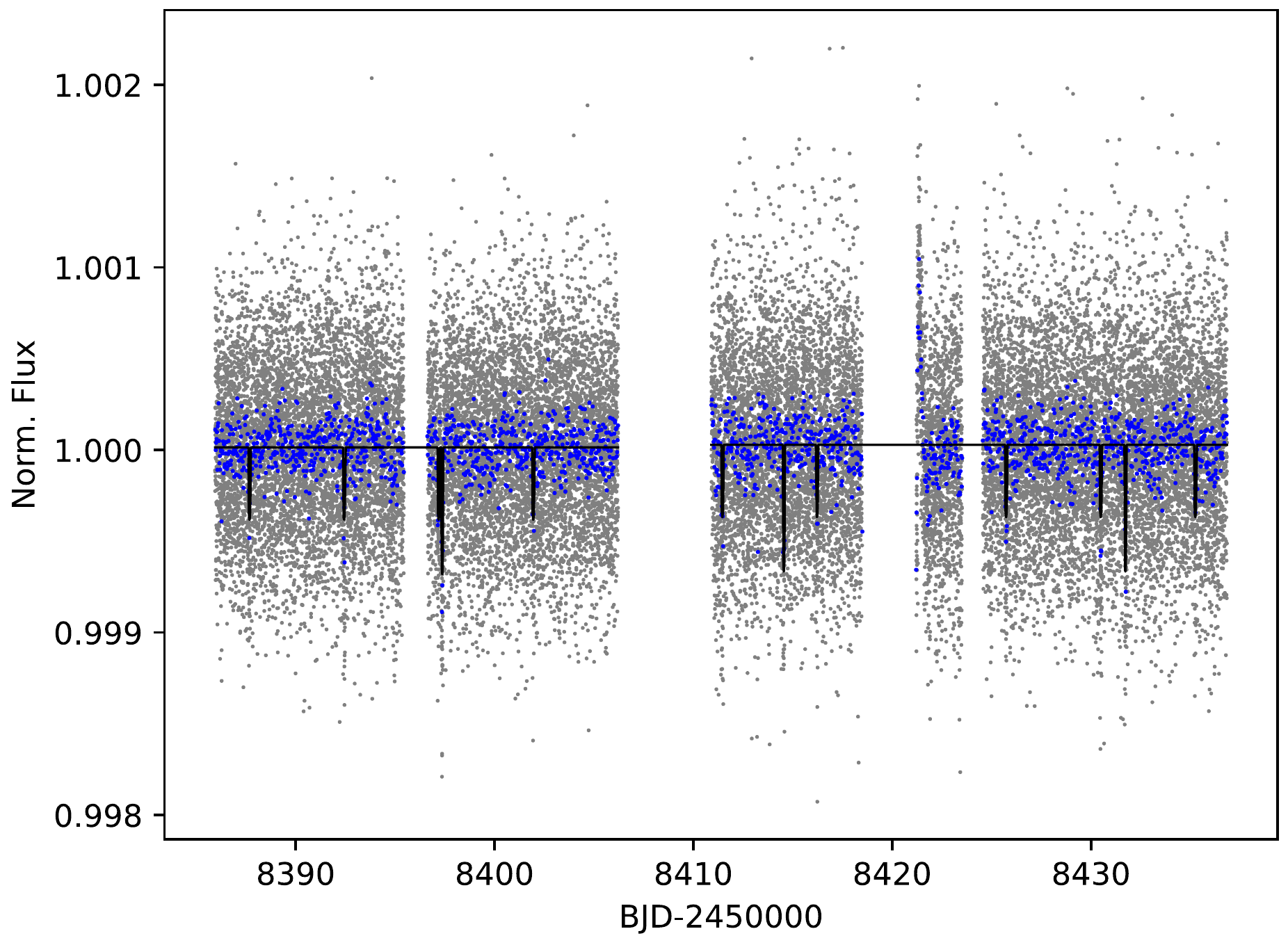}
        \includegraphics[width=6cm]{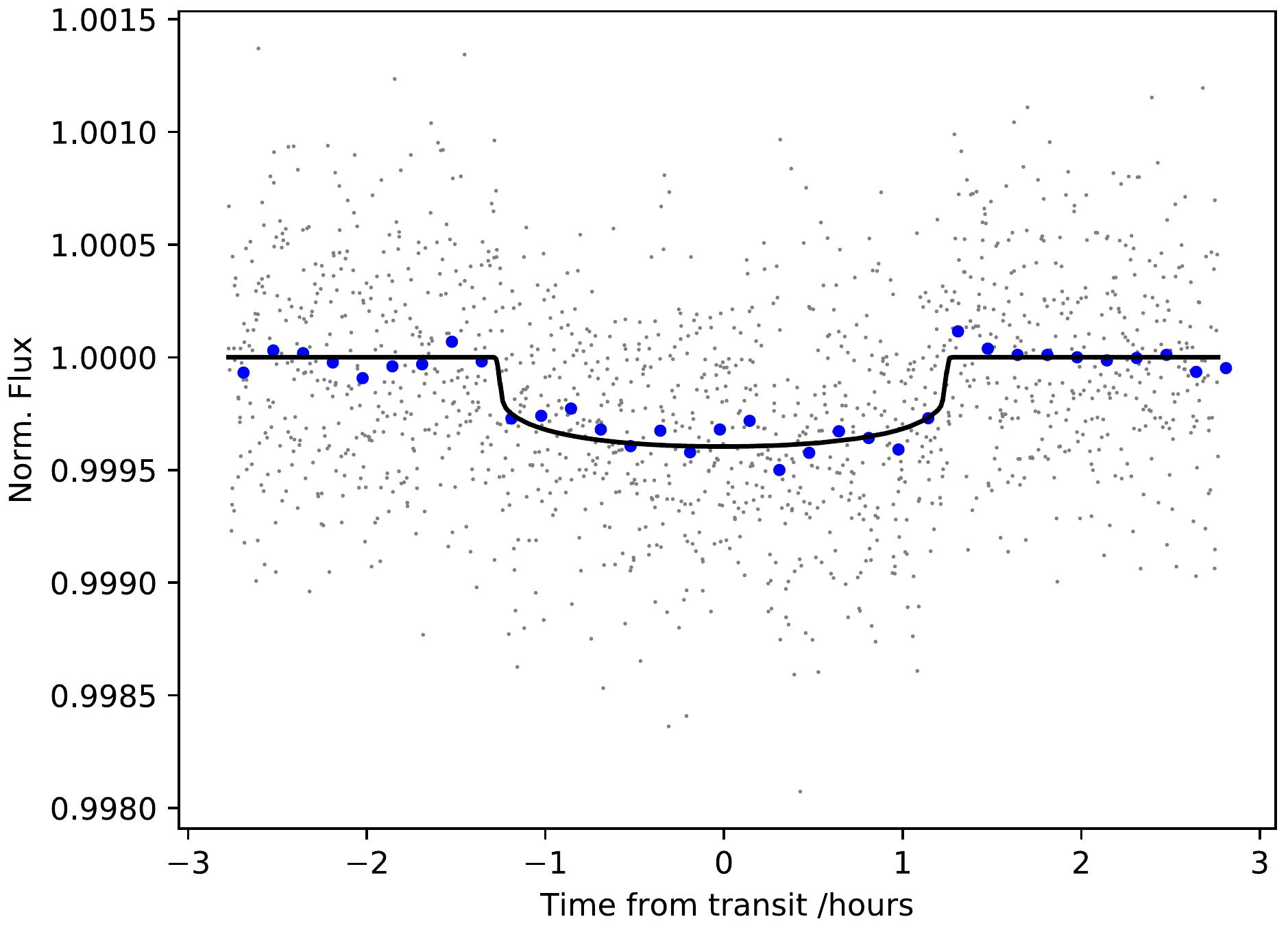}
        \includegraphics[width=6cm]{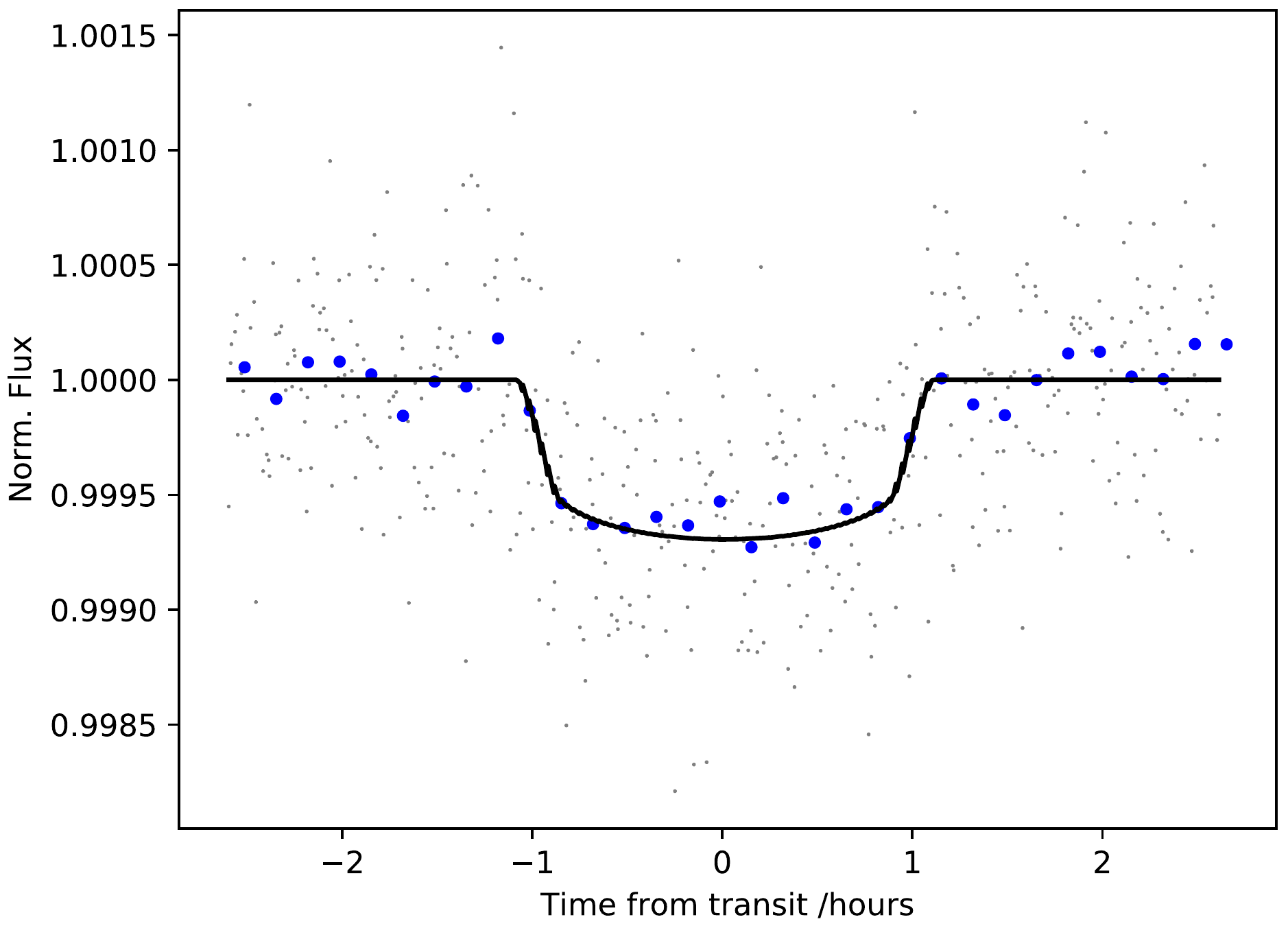}
        \caption{\emph{Left:} Full PDCSAP light curve from \emph{TESS} after removing the smoothing model described in Sect.\,\ref{sec:analysis_TESS}. The grey points represent the raw measurements, while the blue dots are the same data binned over 10 minutes. The model showing the best fit to the light curve with two planets corresponding to TOI-402.01 and TOI-402.02 is shown as the black curve.
                      \emph{Middle:} Light curve phase-folded in phase to highlight the transit of TOI-402.01 and the best-fit model.
                      \emph{Right:} As in the middle plot, but for TOI-402.02.}
        \label{fig:0}      
\end{figure*}

We present the results of the stellar fit in Table \ref{tab:0} and Fig.\,\ref{fig:0}, and the system parameters derived from \emph{TESS} photometry in Table \ref{tab:phot_par}. We note that to test that the smoothing method removing long-term variation in the light curve due to stellar activity was not perturbing the transit depth of both planets, we ran EXOFASTv2 with and without this smoothing. Both runs gave transit depths that differ by no more than 17 ppm, that is, less than the transit depth uncertainties, implying that the smoothing does not bias the final planetary radius. We also compared the results of analyses with eccentricity fixed at zero and using the value and uncertainties given by the RV analysis as the prior. The results were entirely consistent between the two and therefore we present the results of the analysis with fixed eccentricity.

Comparing our results for the time of transit, orbital period, and transit depth with the values found by the Science Processing Operations Center (SPOC) data validation (DV) analysis\footnote{ those values can be found on ExoFOP (\url{https://exofop.ipac.caltech.edu/tess/target.php?id=120896927}). We note however that the ExoFOP value for the period of TOI-402.02 is coming from somewhere else as the SPOC DV state 17.18187$\pm$0.00304 days and not 17.2148 days. To perform the comparison with our results, we used the results from the SPOC DV analysis}, we find that our values are compatible with the SPOC DV analysis within 1$\sigma$ for the time of transit and orbital period of both planets. The transit depths of both planets are compatible to within 2$\sigma$. Our re-analysis of \emph{TESS} photometry gives in all cases a better precision than the values found on ExoFOP.

%
\begin{table*}[]
     \caption{Parameters for the TOI-402 planetary system derived from our \emph{TESS} photometry analysis using EXOFASTv2.}
    \centering
    \begin{tabular}{llcc}
\hline\hline
Symbol & Parameter (Unit) & TOI-402.01 & TOI-402.02 \\
\hline
$P$ &Period (days) &$4.75642^{+0.00021}_{-0.00020}$&$17.1784^{+0.0016}_{-0.0015}$\\
$R_P$ &Radius (\Rjup) &$0.1516^{+0.0055}_{-0.0053}$ (3.6\%)&$0.2250^{+0.0095}_{-0.0091}$ (4.2\%)\\
$R_P$ &Radius (\Rearth) &$1.699^{+0.062}_{-0.059}$&$2.522^{+0.106}_{-0.102}$\\
$T_c$ &Optimal conjunction Time (BJD$_{TDB}$-2400000) &$58411.46202^{+0.00071}_{-0.00083}$&$58414.5500\pm0.0012$\\
$a$ &Semi-major axis (AU) &$0.05245^{+0.00085}_{-0.00072}$&$0.1235^{+0.0020}_{-0.0017}$\\
$i$ &Inclination (Degrees) &$88.36^{+0.36}_{-0.30}$ (0.4\%) & $88.413^{+0.055}_{-0.054}$ (0.06\%)\\
$T_{eq}$ &Equilibrium temperature (K) &$1006^{+13}_{-14}$&$656.1^{+8.4}_{-9.0}$\\
$R_P/R_*$ &Radius of planet in stellar radii  &$0.01856\pm0.00048$&$0.02754^{+0.00091}_{-0.00090}$\\
$a/R_*$ &Semi-major axis in stellar radii  &$13.43^{+0.35}_{-0.33}$&$31.63^{+0.82}_{-0.77}$\\
$\delta$ & Transit depth &$0.000344\pm0.000018$&$0.000758^{+0.000051}_{-0.000049}$\\ 
$T_{14}$ &Total transit duration (days) &$0.1064^{+0.0015}_{-0.0016}$&$0.0930\pm0.0023$\\
$T_{FWHM}$ &FWHM transit duration (days) &$0.1041^{+0.0015}_{-0.0017}$&$0.0830\pm0.0027$\\
$b$ &Transit Impact parameter  &$0.386^{+0.061}_{-0.078}$&$0.876^{+0.010}_{-0.011}$\\
\hline
\multicolumn{2}{l}{Wavelength considered:}&\multicolumn{2}{c}{TESS band-pass}\\
$u_{1}$ &linear limb-darkening coeff  &$0.389^{+0.037}_{-0.036}$\\
$u_{2}$ &quadratic limb-darkening coeff  &$0.220\pm0.036$\\
\hline
    \end{tabular}
    \label{tab:phot_par}
\end{table*}

\subsection{Stellar rotation and activity}
\label{sec:stellar_rot}

To probe the activity of HD\,15337, we analysed the calcium \logrhk\,time-series derived from the HARPS data-reduction software. The \logrhk\,is an observable derived from the S-index, which is
itself a measurement of the emission in the core of the \ion{Ca}{II} H and K lines normalised to nearby pseudo-continuum \citep[][]{Vaughan-1978}. Because the level of those pseudo-continuum is dependant on spectral type,
the S-index value cannot be compared between different stars without correcting for the stellar bolometric luminosity, which is performed in the \logrhk\,\citep[][]{Noyes-1984}. We note that the \logrhk\,derived
from the HARPS pipeline has been calibrated to deliver similar result to those of the band-pass photometer of the Mount-Wilson project \citep[][]{Vaughan-1978}. 

\begin{figure}[h]
        \centering
        \includegraphics[width=9cm]{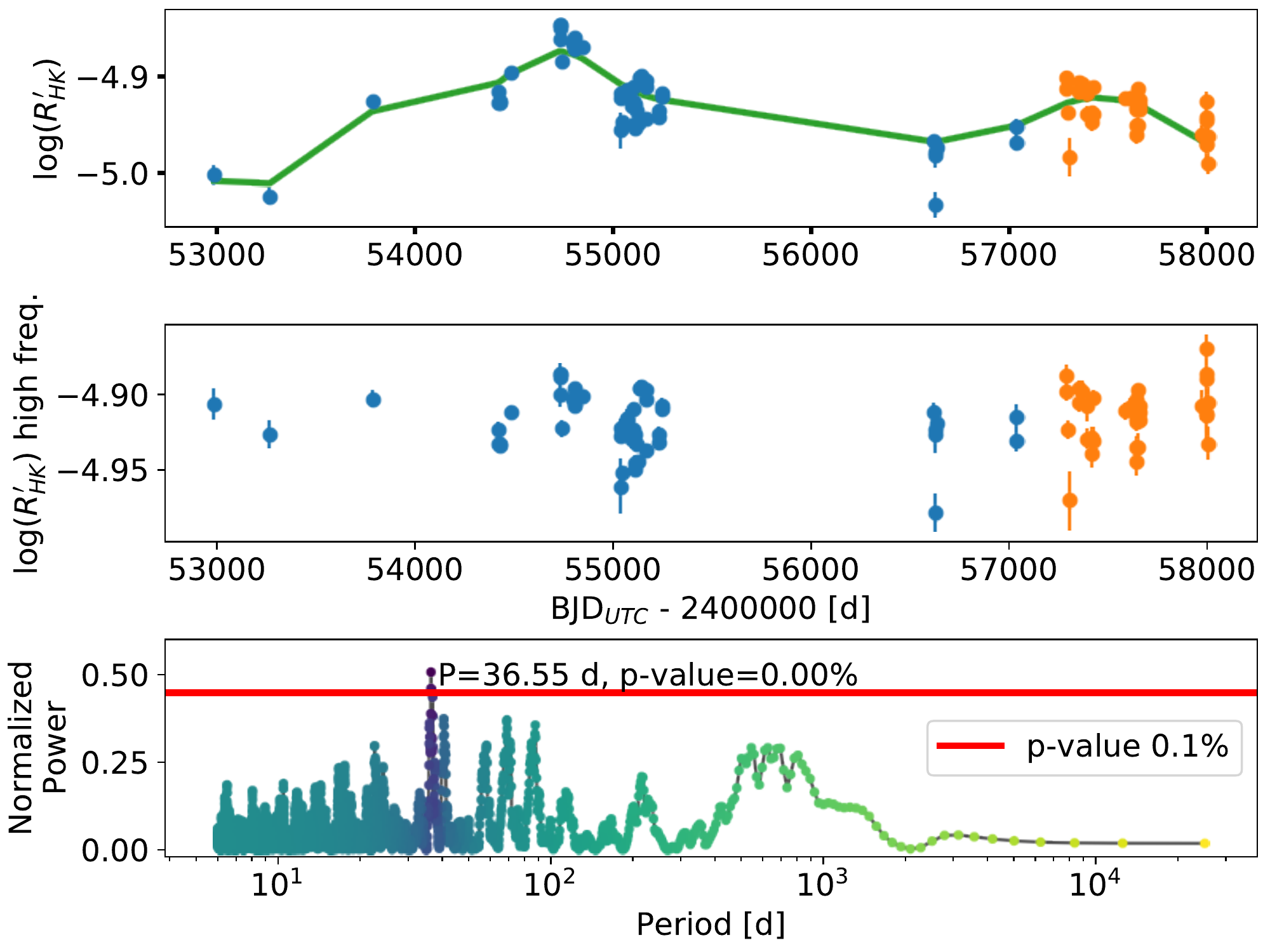}
        \caption{\logrhk\,time-series analysis for HD\,15337.
                      \emph{Top:} Raw time-series and low-pass filter shown as the green line. Blue points represent the data obtained before the change of fibre on June 1, 2015, and orange points show the data obtained afterwards. 
                      \emph{Middle:} \logrhk\,residual time-series after removing the contribution of the low-pass filter. We refer to this time-series as the high-frequency \logrhk\,in the rest of the paper.
                       \emph{Bottom:} Periodogram of the \logrhk\,residuals with the 0.1\% $p-$value shown as the horizontal red line. A significant peak appears at 36.5 days corresponding to the stellar 
                        rotation period of HD\,15337.}
        \label{fig:1}      
\end{figure}

As can be seen in the top panel of Fig. \ref{fig:1}, HD\,15337 presents a magnetic cycle. When fitting a sinusoidal signal to the \logrhk, we find a period of 7 years, that is, shorter than the solar magnetic cycle. However, we must note that this period is not well constrained. The amplitude of this cycle is 
comparable to the Sun with \logrhk\,values ranging from $-$5.0 to $-$4.85 with a mean value of $-$4.91. To see if we can detect any activity on the rotational period timescale, we applied a low-pass filter to the \logrhk\, using the algorithm
described in \citet{Rybicki-1995} and selecting a cutting frequency at 1000 days$^{-1}$. By removing this long-timescale variation from the raw \logrhk, we obtained the time-series seen in the middle panel of Fig.\ref{fig:1}, which we refer to as the
high-frequency \logrhk\, time-series in the rest of the paper.
A generalized Lomb-Scargle periodogram \citep[GLS,][]{Zechmeister-2009} of this time-series is seen in the bottom of the same figure. A significant peak, with a $p$-value smaller than 0.01\% and a period of 36.55 days, reveals
the stellar rotational period of the star ($P_\mathrm{rot}$). Using the relation from \citet{Mamajek-2008} to estimate rotation periods from the colour index B$-$V and the mean \logrhk\, level, we predict a rotation period of $43\pm5$ days, which is compatible with our measurement
within 1.5$\sigma$. We are therefore confident that 36.55 days is the stellar rotation period measured at the average latitude where active regions are present.

From this analysis of stellar activity, we expect the RVs to be affected on timescales similar to the stellar rotation period, and we also expect a small long-term contribution with an amplitude of $\sim4$\,\ms due to the stellar magnetic cycle \citep[][]{Lovis-2011b}. We note that the RV signal induced by TOI-402.02 with an orbital period of 17.17 days might be significantly affected by activity as it is close to half the stellar rotation period. Indeed, we know from simulations 
\citep[e.g.][]{Boisse-2011} and observations that activity signals appear at the stellar rotational period and respective harmonics ($P_\mathrm{rot}$, $P_\mathrm{rot}/2$, $P_\mathrm{rot}/3$, $P_\mathrm{rot}/4$).

\subsection{Radial-velocity analysis}
\label{sec:RV_analysis}

\subsubsection{Preliminary periodogram analysis}
\label{sec:preliminary_analysis}

\begin{figure}[h]
        \centering
        \includegraphics[width=8cm]{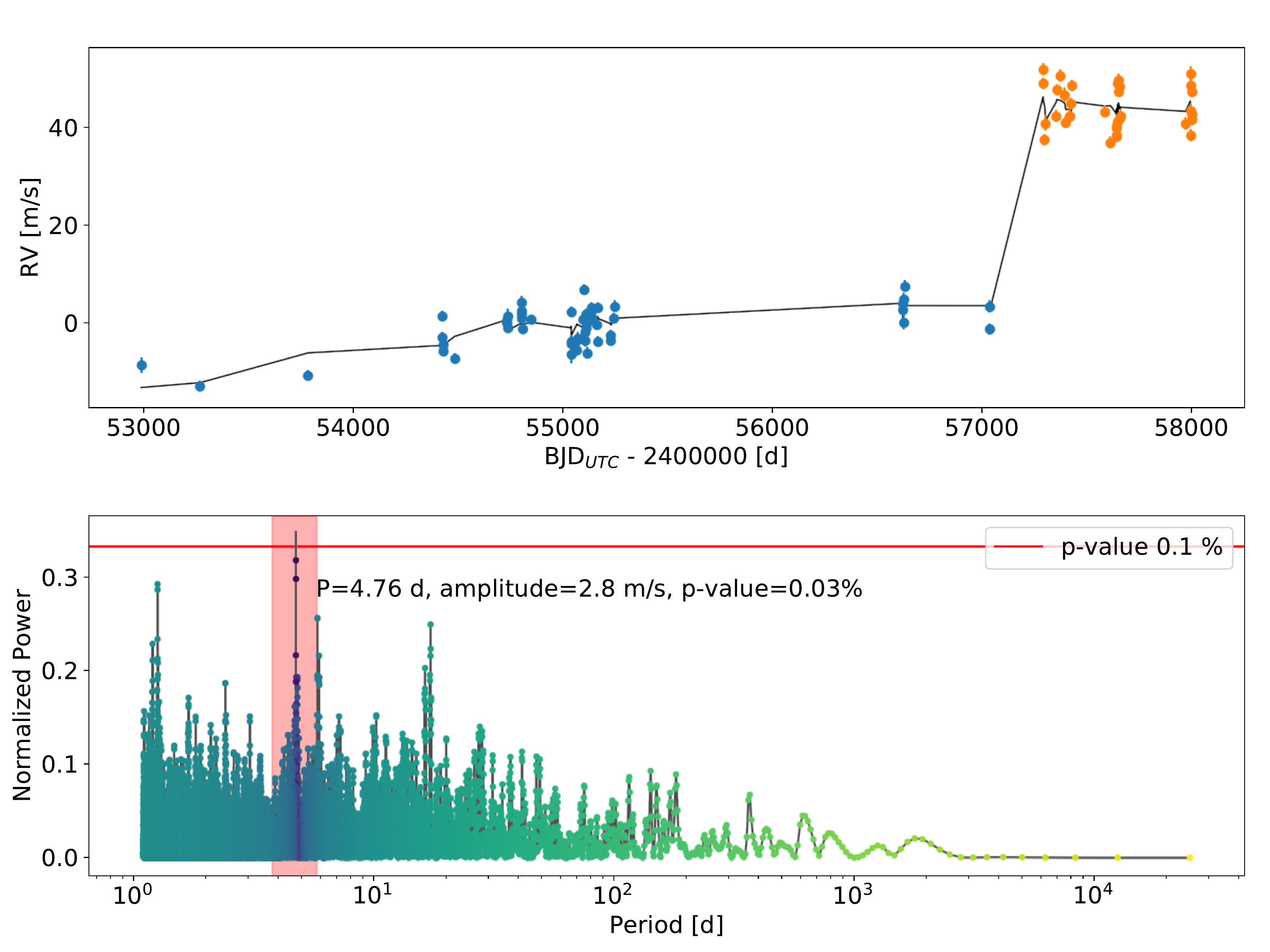}
        \includegraphics[width=8cm]{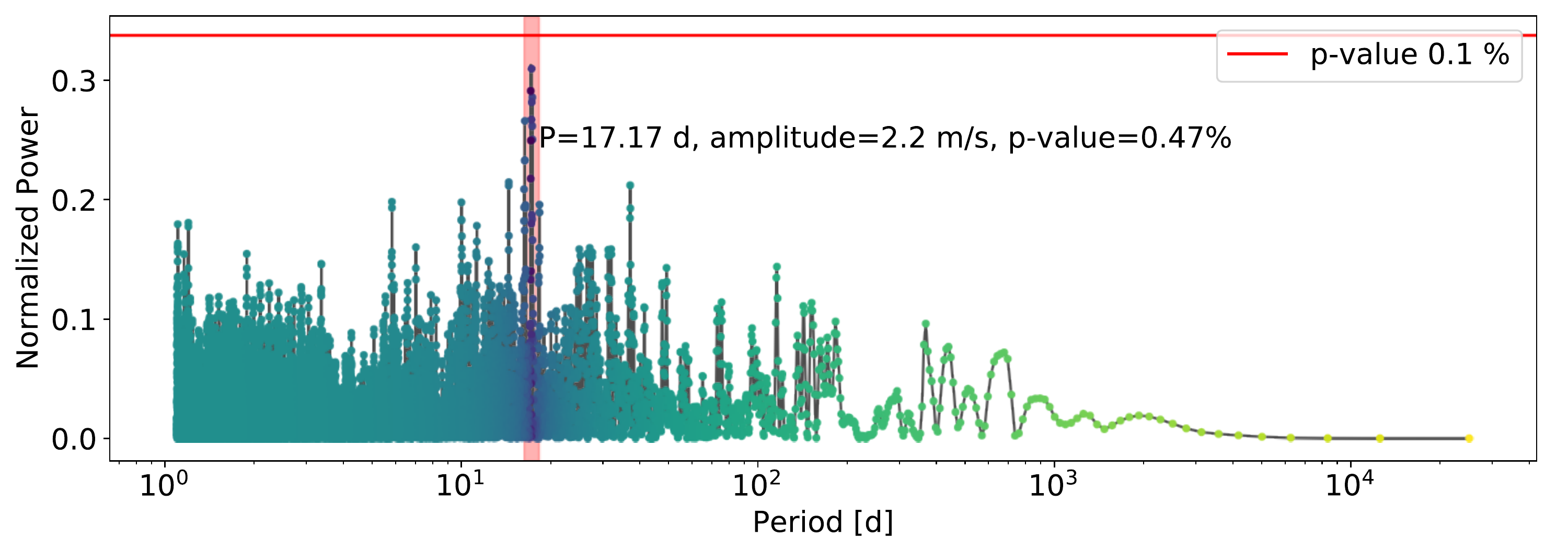}
        \caption{\emph{Top:} Fit composed of an offset to account for the change of fibres on June 1, 2015, a second-order polynomial drift, a linear dependence with the low-pass, and the high-frequency \logrhk\, times series, over-plotted on the raw RVs of HD\,15337
                       \emph{Middle:} GLS periodogram of the RV residuals after fitting the model shown in the top panel. The planetary signal of TOI-402.01 is significant at 4.76 days.
                       \emph{Bottom:} GLS periodogram of the RV residuals after fitting the model shown in the top panel in addition to the signal of TOI-402.01. The signal of TOI-402.02 at 17.17 days appears in this periodogram, however with a 
                       moderate significance of 0.5\%.}
        \label{fig:2}      
\end{figure}

To search for the signals of TOI-402.01 and TOI-402.02, the two planet candidates that \emph{TESS} found orbiting HD\,15337, we analysed the HARPS RVs using a periodogram approach. 
To consider stellar activity and possible long-term drift in the data due to long-period planets or stellar companions, we first fitted a model 
including a linear dependence with the low-pass filtered \logrhk\,time-series (see Sect.\,\ref{sec:stellar_rot} and top panel of Fig.\,\ref{fig:1}), the high-frequency \logrhk\,time-series (see Sect.\,\ref{sec:stellar_rot} and the middle panel of Fig\,\ref{fig:1}),
and a second-order polynomial. In addition, we added an offset to adjust the change of HARPS fibres on June 1, 2015. The model that best fits the data using a linear least square can be seen in the top panel of 
Fig.\,\ref{fig:2}. The middle panel of the same figure shows the GLS periodogram of the RV residuals, which highlight a significant signal corresponding to TOI-402.01 with a $p-$value of 0.03\%, 
an estimated amplitude of 2.8\,\ms , and a period of 4.76 days, fully compatible with the period found by \emph{TESS}

In a second step, we added to the previous model a Keplerian model to account for the signal found at 4.76 days. Before performing a least square on this new model which is non-linear, we require good initial guesses
for all the parameters. For the offset, polynomial drift, and linear dependence with activity, we used as initial conditions the best values found in the precedent fit. For the Keplerian model, the period
is fixed to the value found in the periodogram, in this case 4.76 days. The eccentricity and mean anomaly are estimated using the amplitude and phase of the signals found at 4.76 days and half of this period, as described 
in \citet{Delisle-2016}. The initial guess for the amplitude of the Keplerian signal is obtained by fixing the period, the eccentricity and mean anomaly to the values guessed above and by performing a linear least square on the RV residuals
of the previous fit. We then use the scipy.optimize least\_squares algorithm to find the best parameters for this new model including the TOI-402.01 Keplerian signal.
The GLS periodogram of the RV residuals of this new fit can be found in the bottom panel of Fig.\,\ref{fig:2}. We see in this periodogram the signal of TOI-402.02 with a $p-$value of 0.47\%, an amplitude of 2.2\,\ms , and a period of 17.17 days.

\begin{figure*}[h]
        \centering
        \includegraphics[width=14cm]{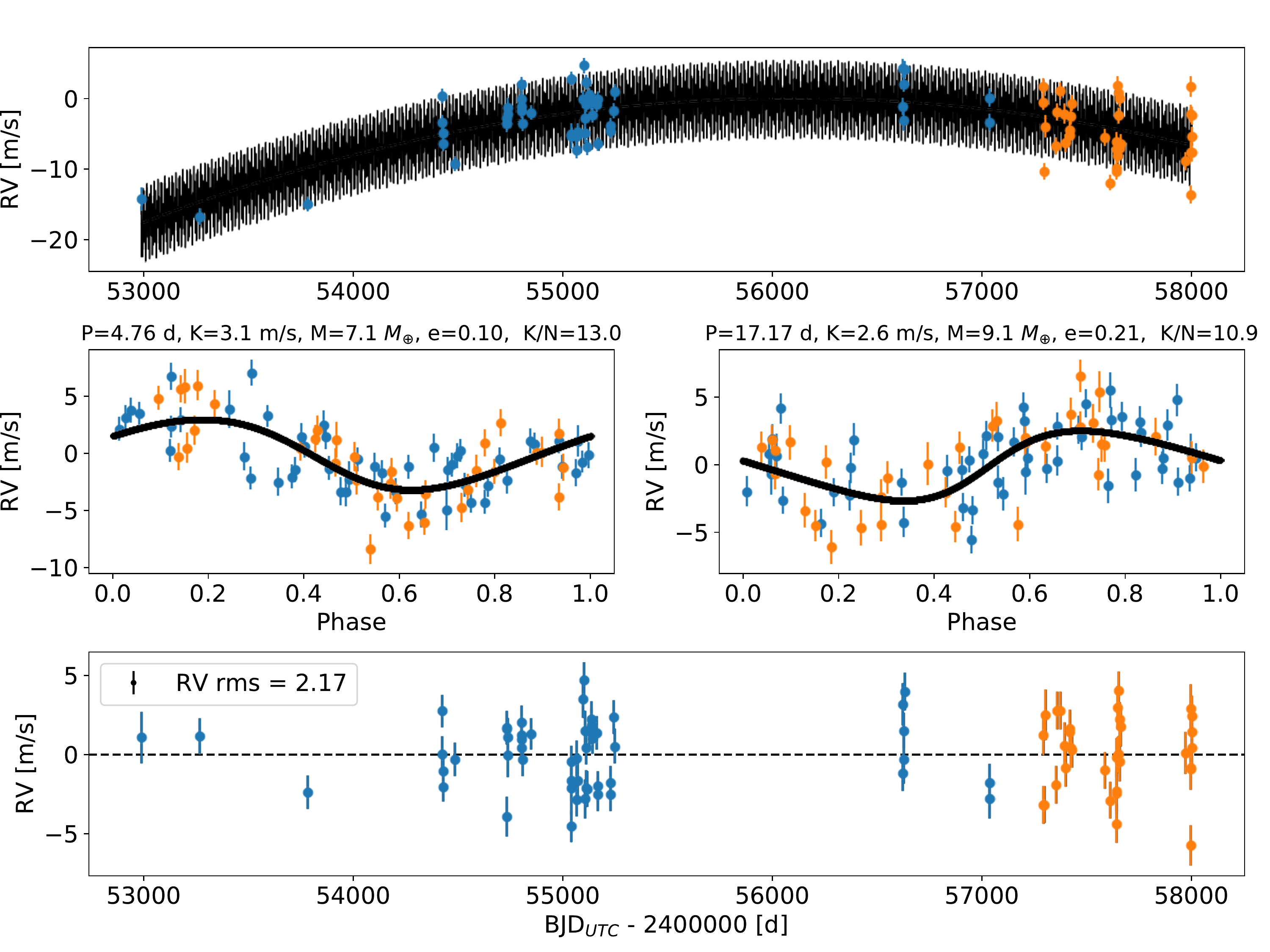}
        \includegraphics[width=14cm]{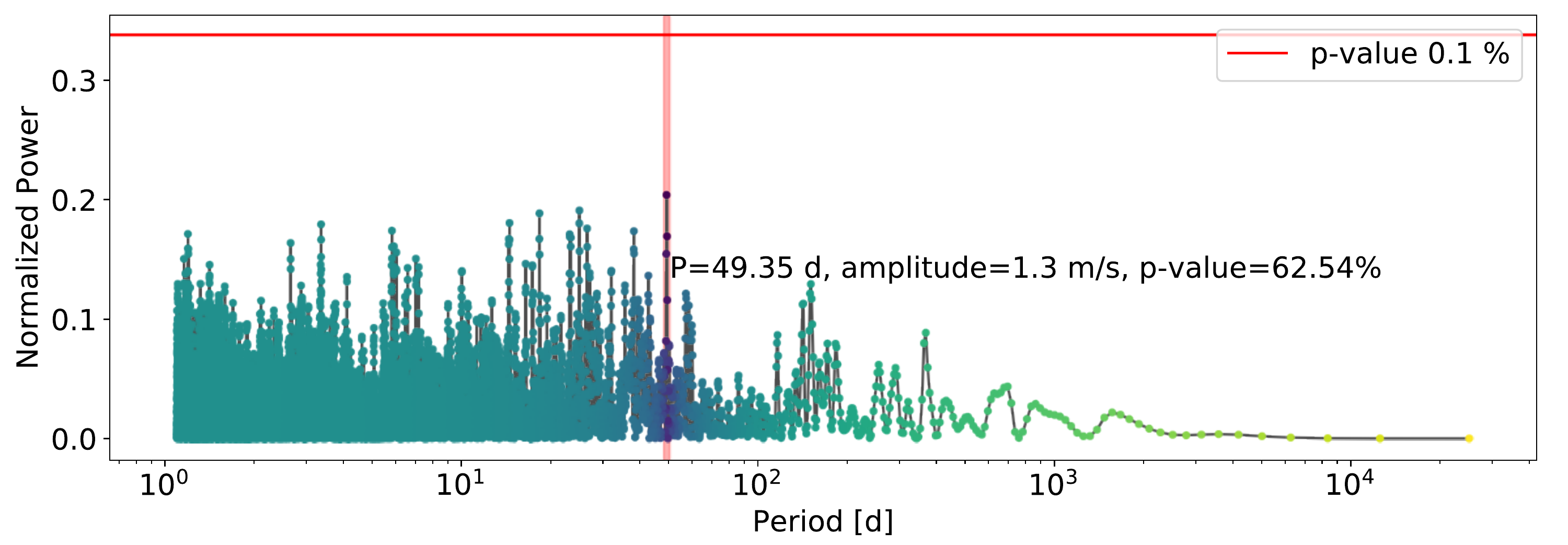}
        \caption{\emph{Top:} Best-fit model of our preliminary RV analysis over-plotted on the RV data of HD15337 after removing the instrumental offset due to the change of the fibres on June 1, 2015. 
                     \emph{Second row:} Phase plot showing the RV signals of TOI-402.01 (\emph{left}) and TOI-402.02 (\emph{right}). The best-fit model is shown in black, and the orbital parameters and mass of the detected 
                     planets can be found in the title of the two subplots.
                     \emph{Third row:} RV residuals after fitting our best model. The rms of the residuals of 2.17\,\ms is significantly higher than the average photon noise error of 0.9\,\ms. 
                     \emph{Bottom:} GLS periodogram of the RV residuals showing no significant signals. Although no signal is seen, the fact that the rms of the residuals is much higher than 
                     the photon noise errors indicates that there is still significant activity signal in the residuals.
                     }
        \label{fig:3}      
\end{figure*}

After finding this second signal, we adopted a new model including the original one plus two Keplerian signals. We found the initial conditions for the Keplerian parameters in the same way as described above and performed a new
least square minimisation. We show the best-fit model in the top panel of Fig.\,\ref{fig:3}. In the two second-row panels, we show the phase curves for the two Keplerian signals corresponding to TOI-402.01 and TOI-402.02.
The two last panels of the same figure show the RV residuals and their corresponding GLS periodogram. It is clear that no other significant signals are found in the data.

Our preliminary modelling of the RVs of HD\,15337 reveals the presence of two super-Earths, with masses of 7.1 and 9.1\,\Mearth, periods of 4.76 and 17.17 days, and eccentricities of 0.10 and 0.21 
for TOI-402.01 and 402.02, respectively. 
The RV residuals after removing our best fit still have a RV rms of 2.17\,\ms, much higher than the precision of the HARPS spectrograph, below 1\,\ms, and the average photon noise RV error 
of the HD\,15337 data, 0.9\,\ms. Therefore, although no significant coherent signal is left in the RV residuals, stellar signal is not fully mitigated. 
This motivates the use in the following section of a more complex model to account for the correlated noise induced by stellar activity.

We note that we tried to remove from our model the linear dependence with the short-term \logrhk\ in order to analyse the impact of stellar activity on the derived planetary solutions. The two planets can still be found, 
although with larger $p-$values. The stellar rotation period is more significant in the RV residuals of the full model with a $p-$value of 3.44\%. Finally, the eccentricities of TOI-402.01 and 402.02
are larger, 0.29 and 0.42, respectively, which is uncommon for small planets on short-period orbits. This analysis of the RVs of HD\,15337, without considering short-term stellar activity, shows that
the planetary signals are perturbed by stellar activity, and therefore it should be taken into consideration to obtain robust orbital solutions for TOI-402.01 and TOI-402.02.
 
This preliminary analysis of the RV data of TOI-402 confirms the two planetary candidates detected by \emph{TESS}. Although the signal for TOI-402.01 is strong and the RV data alone are sufficient 
to confirm the planet, the second signal at 17.17 days has a $p-$value of 0.47\% and is not at the 0.1\% threshold generally required to claim a confident detection. Of course,
the \emph{a\,priori} knowledge that a 17.17-day transiting signal is found in \emph{TESS} data validates the  planetary nature of this signal.

\subsubsection{Mitigating stellar activity in \logrhk\, using a Gaussian process regression}
\label{sec:GP_training_Rhk}

When RV data are significantly affected by stellar activity, it is common to use either a Gaussian process regression model \citep[GP, e.g.][]{Haywood-2014, Grunblatt:2015aa, Rajpaul-2015, Faria-2016a, 
Jones:2017aa, Damasso:2018aa} or a moving average regression model \citep[e.g.][]{Tuomi-2013, Tuomi-2014a, Feng:2016aa, Feng:2017aa} to account for the correlated signal induced by stellar activity. 
These two techniques have been shown to be the most efficient at mitigating stellar activity while looking for planetary signals \citep[][]{Dumusque:2017aa}.

In this section, we use a GP regression model to account for stellar activity. It is common in the case of good photometry to first train the GP on photometry to obtain good constraints on its hyper-parameters
and then to use those constraints as priors for a GP regression model applied to the RV data set. This is possible with \emph{Kepler}, due to photometric time-series spanning many stellar rotation periods. 
However, in the case of HD\,15337,  \emph{TESS} obtained a photometric time-series spanning a total of only 51 days, representing less than two full rotational periods of the star, estimated to be 36.55 days (see Sect.\,\ref{sec:stellar_rot}).
In addition, the \emph{TESS} photometric time-series in not contemporaneous with the RV measurements and therefore the covariance observed in photometry might be different from the one present in the RV data.
We therefore decided to use the \logrhk\,time-series extracted from HARPS spectra. The \logrhk\,time-series is affected by a long-term magnetic cycle. Here, we want to model with the GP regression only the
rotational modulation induced by stellar activity, and therefore we analysed the high-frequency \logrhk\,time-series discussed in Sect.\,\ref{sec:stellar_rot} and shown in the middle panel of Fig.\,\ref{fig:1}.

For our GP regression, we tested two different kernels. First a simple squared exponential (SE) kernel, which takes into account correlations between points inversely proportional to distance, and a more complex quasi-periodic (QP) kernel which adds a periodic component to it and therefore is often used to model the semi-periodic signals induced by stellar activity that decorrelates with time due to active regions evolving on the surface of a star \citep[e.g.][]{Pont-2013, Haywood-2014, 
Lopez-Morales:2016aa, Cloutier:2017aa, Astudillo-Defru:2017aa, Bonfils:2018aa, Diaz:2019aa}. These two kernels result in covariance matrices of the following forms:
\begin{eqnarray}\label{eq:rhk}
\mathcal{K}_{ij,\,SE} &=& A_{\log(R^{\prime}_\mathrm{HK})}^2\,\exp\left[-\frac{(t_i-t_j)^2}{2\tau^2}\right] + \delta_{ij}(\sigma_i^2 + s_{\log(R^{\prime}_\mathrm{HK})}^2),\\
\mathcal{K}_{ij,\,QP} &=& A_{\log(R^{\prime}_\mathrm{HK})}^2\,\exp\left[-\frac{(t_i-t_j)^2}{2\tau^2} - \Gamma^2\,\sin^2\left(\frac{\pi}{P_\mathrm{rot}}|t_i - t_j|\right)\right]\\
                         &+& \delta_{ij}(\sigma_i^2 + s_{\log(R^{\prime}_\mathrm{HK})}^2),  \nonumber
\end{eqnarray}
where $A_{\log(R^{\prime}_\mathrm{HK})}$ is the amplitude of the GP regression modelling the high-frequency \logrhk, $\tau$ is the de-correlation timescale, $P_\mathrm{rot}$ is the stellar rotational period, and $\Gamma$ 
is the scale of the correlation \citep[e.g.][]{Rasmussen-2006}. The last term in both kernels takes into account white noise in the data, which has for variance the sum between the \logrhk\,photon noise variance $\sigma_i^2$ 
and a parameter $s_{\log(R^{\prime}_\mathrm{HK})}^2$ that accounts for additional white noise. In addition to this covariance, a constant will be adjusted to the high-frequency \logrhk\,time-series to adjust for the mean. 

We used the \emph{george} package \citep[][]{Foreman-Mackey-2015b} for building the covariance and predicting the GP regression on the \logrhk\,time-series and the \emph{emcee} package \citep[][]{Foreman-Mackey-2013} 
for sampling, using the affine invariant MCMC ensemble sampler \citep[][]{Goodman:2010aa} and the marginalised posterior probability density functions (PDF) of the model parameters with the goal of 
maximising a defined likelihood function. In the case of a covariance matrix where elements outside of the diagonal are not null, as is the case here with our GP quasi-periodic kernel, the log-likelihood 
$ln\,\mathcal{L}(\boldsymbol{r}|\Theta)$ takes the form:
\begin{equation}\label{eq:likelihood}
ln\,\mathcal{L}(\boldsymbol{r}|\Theta) = -\frac{1}{2}\left(\boldsymbol{r}^{\emph{T}}\mathcal{K}^{-1}\boldsymbol{r} + \ln\,{\mathrm{det}\,\mathcal{K}} + N\,\ln\,2\pi \right).
\end{equation}

In this equation, $\boldsymbol{r}$ is the vector of residuals, in this case the high-frequency \logrhk\,time-series minus the constant parameter to adjust the mean, $\Theta$ is a list of all model parameters,
$N$ is the number of observations, and `det' stands for determinant.

Before performing any analysis, we re-centred the high-frequency \logrhk\,time-series around its mean.
We tried first to train the GP regression with either the SE or QP kernels by letting all the parameters free to vary with loose priors, but due to the lack of data and the bad sampling of consecutive stellar rotation periods it was not possible to constrain $\tau$ in the case of the SE kernel, and $\tau$, $P_\mathrm{rot}$, and $\Gamma$ in the QP kernel. We therefore decided to fix those parameters.

By the analysis of activity performed in Sect.\,\ref{sec:stellar_rot}, it is clear that the stellar rotation period is somewhere close to 36.55 days. We therefore fixed $P_\mathrm{rot}$ to 
this value. The parameter $\tau$ corresponds to the average lifetime of active regions. We first tried to use the relation found in \citet{Giles:2017aa} that uses the scatter in photometry and the stellar effective temperature
to estimate the lifetime of spots. With a \emph{TESS} light-curve rms of 0.49 mmag and a stellar effective temperature of 5131\,$K$, Eq. 8 in \citet{Giles:2017aa} gives a spot lifetime of 24.9 days
\footnote{We note that in Eq. 8 of \citet{Giles:2017aa} , the authors use for rms the 2$\sigma$ range of the rms, therefore twice the conventional rms.}, much shorter
than the stellar rotation period. This is not expected and could be due to several problems. The relation found in \citet{Giles:2017aa} is calibrated for stars with rotation periods smaller than 20 days, thus 
more active than HD\,15337. In addition, it was based on \emph{Kepler} photometry, which is performed in a bluer filter than \emph{TESS}, and therefore should be more affected by stellar activity due to spot contrast 
increasing towards the blue. On the Sun, the average lifetime of main active regions is about two stellar rotations \citep[][]{Howard-2000}. We therefore decided to fix $\tau$ at twice $P_\mathrm{rot}$, that is, 73.1 days.
Finally, we needed to set the value for $\Gamma$. Looking at the literature, we found  values 
for $\Gamma$ of 2.2, 1.2, 2.08, and 2.0  in  \citet{Lopez-Morales:2016aa}, \citet{Cloutier:2017aa}, \citet{Damasso:2017aa}, and \citet{Dittmann:2017aa} respectively. We therefore decided to use a fixed value of 2.0 here as in any case, the exact value of this parameter is not critical.
%

Before sampling the posterior PDF of both GP models with the kernels described above using the \emph{emcee} MCMC, we first initialised the different parameters to the values listed in Table\,\ref{tab:1}. 
We then used the scipy \emph{minimize} algorithm to obtain an estimate of the maximum \emph{a posteriori} (MAP) solution for the different parameters.
Subsequently, we selected the priors for each parameter as described in Table\,\ref{tab:1} and performed ten independent MCMC runs starting with the MAP solution, each of them having
four times more walkers than the number of parameters, that is, 12 here. For each run, a tiny amount of noise ($10^{-8}$) was added to the MAP solution before starting sampling.
Each run consists of a burn-in phase of 5,000 iterations, followed by a production phase of 100,000 iterations. We note that after the burn-in phase, the highest 
likelihood solution is kept and injected, with a tiny amount of noise ($10^{-8}$), as initial condition to the production phase.

To check that our MCMC chains properly converged, we used the GR statistic.
Within an \emph{emcee} run, there are 12 chains per parameter, corresponding to the number of selected walkers; however, those chains are not
independent as they cross-talk with each other \citep{Goodman:2010aa}. We therefore compared chains within the ten independent MCMC runs that we
performed. For each parameter, the GR statistics is computed for all the chains of walkers 1 up to 12 in the ten independent runs. We therefore get 12 GR statistics values 
per parameter. In our case, all the GR statistics values were below 1.01, indicating that our chains properly converged. After checking the autocorrelation timescale for each 
parameter and each walker chain, we kept only one step every 1,000 iterations. We then combined for each parameter the chains from all the corresponding walkers and for the ten independent MCMC runs.

In Figures\,\ref{fig:app1} we show the correlation between the different marginalised posterior PDFs of the model parameters sampled by our MCMC when using a SE kernel. No significant correlation is observed, and the marginalised posterior PDFs follow a well-sampled Gaussian distribution. The marginalised posterior PDFs in the case of the QP kernel are identical. The mean of each marginalised posterior PDF with its 68\% uncertainty for both the SE and QP kernels are given in Table\,\ref{tab:1}. As we can see, the results for the SE and QP kernels are fully compatible within error bars, and therefore in such a case, Occam's Razor principle tell us that the simpler model should be considered. The SE kernel is therefore favoured, and a likely explanation for that is that the data in hand are not well sampled at the stellar rotation period of the star, which therefore prevents a QP kernel from adding a significant improvement.

\begin{figure*}[h]
        \centering
        \includegraphics[width=18cm]{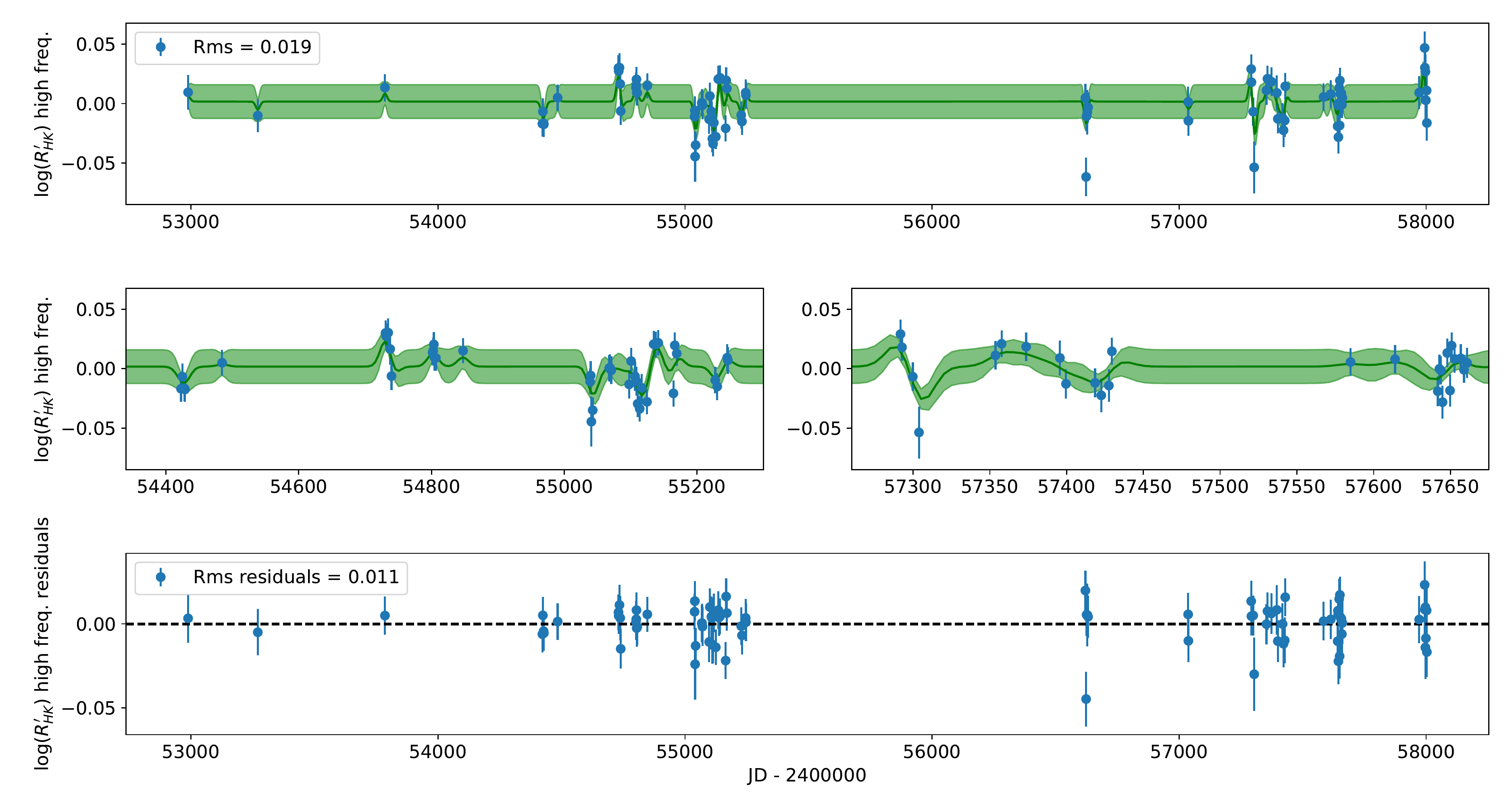}
        \caption{\emph{Top:} Gaussian process regression over-plotted on  the high-frequency \logrhk\,time-series. The rms of those data is 0.019 dex. \emph{Middle:} As in the top panel, however each subplot is a zoom on times where
                                        stellar activity is well sampled by the data. \emph{Bottom:} Residuals of the high-frequency \logrhk\,time-series after removing the GP regression. The rms of those residuals is 0.011 dex.}
        \label{fig:4}      
\end{figure*}

The GP regression model using a SE kernel with the mean values listed in Table\,\ref{tab:1} is over-plotted on the high-frequency \logrhk\,time-series in the top panel of Fig.\,\ref{fig:4}. The two middle panels show the same
result, however zoomed-in to  better highlight the GP regression during times when a significant amount of data is present. Finally, the residuals are shown in the bottom panel. 
The \logrhk\,rms decreases from 0.019 in the high-frequency \logrhk\, time-series to 0.011 in the \logrhk\,residuals, which demonstrates that our GP regression is able to significantly mitigate stellar activity, even though we had to guess a value for the de-correlation timescale of the SE kernel.
\renewcommand{\arraystretch}{1.3}
\begin{table*}[ht]
        \scriptsize
        \caption{Initial parameters, priors, and best solutions for the models fitted to the \logrhk ~and the RVs.}            
        \label{tab:1}     
        \centering                         
        \begin{tabular}{lccccc}       
                \hline\hline
                Parameter description & Parameter & Initial value & Prior & \multicolumn{2}{c}{Mean of marginalised posterior PDFs with 68\% confidence interval errors}\\
                & & & & Squared Exponential kernel & Quasi Periodic kernel\\
                
                \hline
                {\bf Fit to \logrhk} & & & & \\
                \hline  
                \multicolumn{6}{c}{\bf{GP for \logrhk}}\\       
                GP amplitude & $A_{\log(R^{\prime}_{HK})}$ & std(\logrhk) = 0.019 & $\mathcal{U}(-1^{20}, 1^{20})$ & 0.0142$\pm$0.0027 & 0.0142$\pm$0.0028 \\      
                GP de-correlation timescale & $\tau$ & 73.1 & fixed & 73.1 & 73.1 \\    
                GP scale of correlation & $\Gamma$ & 2 & fixed & - & 2 \\
                GP rotation period & $P_{rot}$ & 36.55 & fixed & - & 36.55 \\
                Extra noise & $s_{\log(R^{\prime}_{HK})}$ & 0.01 & $\mathcal{U}(-1^{20}, 1^{20})$ & 0.0100$\pm$0.0017 & 0.0100$\pm$0.0017 \\
                Constant & Cte & 0 & $\mathcal{U}(-1^{20}, 1^{20})$ & 0.0017$\pm$0.0031 & 0.0017$\pm$0.0031 \\
                \\
                \logrhk\,high frequency residuals rms    &  &  &  & 0.011& 0.011\\
                \hline
                {\bf Fit to RV} & & & & \\
                \hline
                \multicolumn{6}{c}{\bf{GP for RV}}\\    
                GP amplitude for RV & $A_{RV}$ & 2.5 & $\mathcal{U}(-1^{20}, 1^{20})$ & 2.3262$\pm$0.4552&\\
                GP de-correlation timescale & $\tau$ & 73.1 & fixed & 73.1& \\    
                
                \multicolumn{6}{c}{\bf{Offset, drift and noise}}\\
                Extra noise (\ms)                                   & $s_{RV}$ & 0.6                        & $\mathcal{U}(-1^{20}, 1^{20})$ & 1.5315$\pm$0.2531& \\
                Offset data before 06.2015 (\ms)         & $off_1$     & -7.48                    & $\mathcal{U}(-1^{20}, 1^{20})$ & -4.0003$\pm$4.3446 & \\
                Offset data after 06.2015 (\ms)            & $off_2$     & 37.97                   & $\mathcal{U}(-1^{20}, 1^{20})$ & 40.6186$\pm$2.0401& \\
                Polynomial linear (\ms days$^{-1}$)      & $poly_1$   & -8.46\,$10^{-3}$   & $\mathcal{U}(-1^{20}, 1^{20})$ & -5.64\,$10^{-3}$$\pm$2.56\,$10^{-3}$& \\
                Polynomial quadratic (\ms days$^{-2}$)   & $poly_2$   & -1.83\,$10^{-6}$   & $\mathcal{U}(-1^{20}, 1^{20})$ & 1.35\,$10^{-6}$$\pm$0.38\,$10^{-6}$& \\
                
                \multicolumn{6}{c}{\bf{TOI-402.01}}\\
                Amplitude (\ms)                                & $K_{1}$ & 3.07                             & $\mathcal{U}(-1^{20}, 1^{20})$ & 3.1079$\pm$0.3515& \\
                Period (days)                                        & $P_{1}$ & 4.75642                        & $\mathcal{N}(4.75642, 0.00021)$ & 4.75630$\pm$0.00016& \\
                Transit time (BJD$_{TDB}$ - 2400000)          & $Tc_{1}$ & 58411.46200               & $\mathcal{N}(58411.46200, 0.00088)$ & 58411.46201$\pm$0.00088& \\
                eccentricity parametrisation            & $e\cos{\omega}_1$ & -3.52\,$10^{-2}$  & $\mathcal{B}\left[0.711, 2.57\right](-1, 1)$ & 0.1291$\pm$0.0845& \\
                eccentricity parametrisation            & $e\sin{\omega}_1$ & 9.78\,$10^{-2}$    & $\mathcal{B}\left[0.711, 2.57\right](-1, 1)$ & 0.0751$\pm$0.0897& \\
                Mass (\Mearth)                               &  &  &  & 7.20$\pm$0.81 (11.3\%)& \\
                Eccentricity                                     &  &  &  & 0.17$\pm$0.09& \\
                Semi-major axis (au)                      &  &  &   & 0.052$\pm$0.001& \\
                
                \multicolumn{6}{c}{\bf{TOI-402.02}}\\
                Amplitude  (\ms)                       & $K_{2}$ & 2.58                             & $\mathcal{U}(-1^{20}, 1^{20})$ & 2.4819$\pm$0.4679& \\
                Period  (days)                                & $P_{2}$ & 17.1784                        & $\mathcal{N}(17.1784, 0.0016)$ & 17.1773$\pm$0.0016& \\
                Transit time (BJD$_{TDB}$ - 2400000)     & $Tc_{2}$ & 58414.5501                 & $\mathcal{N}(58414.5501, 0.0013)$ & 58414.5501$\pm$0.0013& \\
                eccentricity parametrisation     & $e\cos{\omega}_2$ & -2.80\,$10^{-2}$  & $\mathcal{B}\left[0.711, 2.57\right](-1, 1)$ & -0.0493$\pm$0.0569& \\
                eccentricity parametrisation     & $e\sin{\omega}_2$ & -2.07\,$10^{-1}$   & $\mathcal{B}\left[0.711, 2.57\right](-1, 1)$ & -0.1551$\pm$0.1197& \\
                Mass (\Mearth)                         &  &  & & 8.79$\pm$1.68 (19.1\%)& \\
                Eccentricity                               &  &  & & 0.19$\pm$0.10& \\
                Semi-major axis (au)                &  &  &  & 0.123$\pm$0.002& \\
                \\
                RV residuals rms (\ms)                 &  &  &  & 1.36& \\
                \hline
        \end{tabular}
\end{table*}

\subsubsection{Gaussian process and Keplerian model fit to the RVs}
\label{sec:GP_fit_RVs}

In the preceding section, we show that a GP regression of the \logrhk\,time series using a SE or QP kernel gives equivalent results and therefore a SE kernel should be used to model stellar activity effects as less complex. By making the hypothesis that
the covariance due to stellar activity seen in the RVs is similar to the one seen in \logrhk, we use the {\bf same de-correlation timescale} to model stellar activity in the RV time-series using a GP regression with a SE kernel. Only the amplitude and extra white-noise terms
are free to vary and are sampled using an MCMC as the values for those hyper-parameters are different between the \logrhk\,and RV time-series.

Based on the preliminary analysis described in Sect.\,\ref{sec:preliminary_analysis}, the model used here to fit the RV data of HD\,15337 is composed of two offsets to account for the jump induced in RV by the change 
of optical fibres on June 1, 2015, a polynomial drift of second order to account for the long-term trend observed in the data, and two Keplerian functions to account for the signal induced by TOI-402.01 and TOI-402.02.
This model can be summarised as follows:
\begin{eqnarray} \label{eq:RV}
RV_{i} &=& \sum_{j} off_j + \sum_{k} poly_k\,time_i^k \\
           &+& \sum_{l} K_l \left(\cos(\omega_l + \nu_i) + e_l\sin(\omega_l)\right), \nonumber
\end{eqnarray}
where $i$ corresponds to each observation, $j=[1,2]$ corresponds to different instruments, in this case two for data before and after June 1, 2015, $k=[1,2]$ for a second-order polynomial, $l=[1,2]$ for the two planets,
$K_l,\,\omega_l$, and $\,e_l$ are the amplitude, argument of periapsis, and eccentricity of planet $l$ and $\nu_i$ is the true anomaly.
In addition to this model, we use a GP regression with a quasi-periodic kernel to account for the short-term stellar activity present in the RVs. This kernel will be similar to the one described in 
Eq.\,\ref{eq:rhk}, with the exception that the amplitude $A_{\log(R^{\prime}_\mathrm{HK})}$ is replaced by $A_\mathrm{RV}$ and the extra white noise jitter term $s_{\log(R^{\prime}_\mathrm{HK})}$ is replaced by $s_\mathrm{RV}$.
As in Sect.\,\ref{sec:GP_training_Rhk}, we sample  the marginalised posterior PDFs of the model parameters with an MCMC with the goal of maximising the likelihood function
described in Eq.\,\ref{eq:likelihood}. In this equation, $\boldsymbol{r}$, the vector of residuals, is replaced by the RV times-series minus $RV_{i}$, our model for the RVs shown in Eq.\,\ref{eq:RV}.

The offsets and polynomial coefficients are initialised to the best values found during the preliminary RV analysis (see Sect.\,\ref{sec:preliminary_analysis}) and can be found in Table\,\ref{tab:1}. 
The period, de-correlation timescale, and length scale of the GP regression are fixed due to the small amount and bad sampling of the data (see Sect.\,\ref{sec:GP_training_Rhk}) and the 
GP amplitude is initialised to 2.5 m/s. The orbital period and time of transit of the two planets are initialised to the values found by the analysis of the \emph{TESS} light curves performed in Sect.\,\ref{subsec:phot}
and their amplitude, $e\cos{\omega,}$ and $e\sin{\omega}$ are initialised to the values found during the preliminary RV analysis (see Sect.\,\ref{sec:preliminary_analysis}). We note here that we sample
over $e\cos{\omega}$ and $e\sin{\omega}$ rather than $e$ and $\omega$, as this makes the posterior PDF easier to sample.

Regarding priors, these are summarised in Table\,\ref{tab:1}. We used tight normal priors for the period and time of transit of the two planets based on the results of our transit analysis (see Sect.\,\ref{sec:analysis_TESS}) as we 
do not expect the RV to provide more constraints on those parameters. We also choose a Beta distribution with parameters 0.711 and 2.57 as prior for the eccentricity, as advised by \citet{Kipping-2013aa}. All the other priors are chosen uniform.

As explained in details in Sect.\,\ref{sec:GP_training_Rhk}, we first searched for the maximum \emph{a posteriori} (MAP) solution for the different parameters and then performed ten independent MCMC 
runs starting with the MAP solution using \emph{emcee} and selecting four times more walkers than the number of parameters, that is, 64 here. Each run consisted of a burn-in phase of 5,000 iterations 
followed by a production phase of 100,000 iterations. We note that after the burn-in phase, the highest likelihood solution is kept and injected with a tiny amount of noise ($10^{-8}$) as initial condition to the production phase.

To check the convergence of the chains, we used, as in Sect.\,\ref{sec:GP_training_Rhk}, the GR statistics over the ten independent MCMC runs. For the 16 parameters and 64 walkers,
we thus obtain 1024 GR statistics values that were all below 1.01, therefore implying that all our chains converged properly. For each chain, we selected one iteration every 1,000 to remove any
autocorrelation. For each parameter we then combined  the chains from all the corresponding walkers and for the ten independent MCMC runs.

\begin{figure*}[h]
        \centering
        \includegraphics[width=18cm]{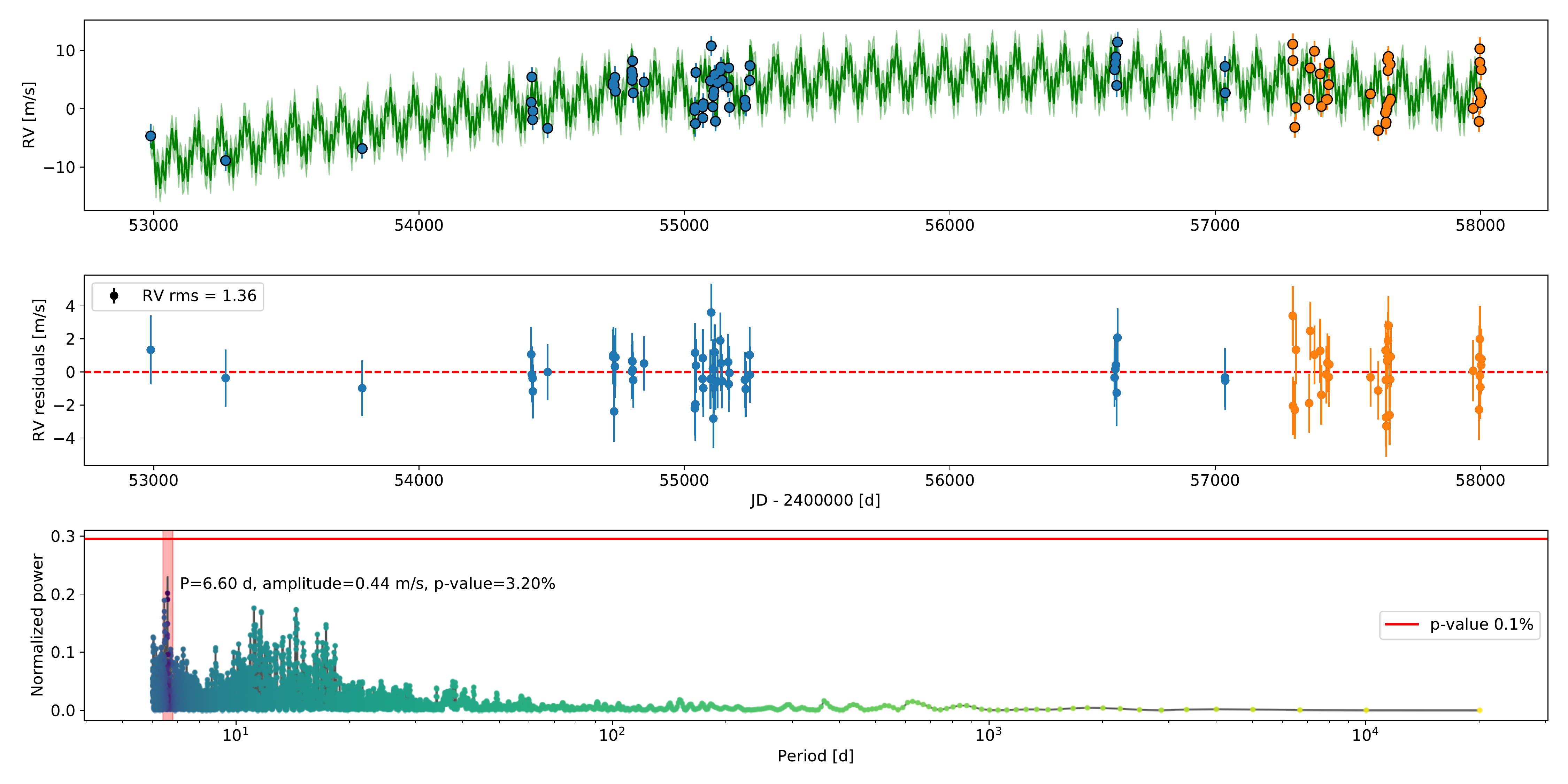}
        \caption{\emph{Top:} Best-fit model using the mean of all marginalised posterior PDFs obtained by the MCMC sampler over-plotted on the raw RV data of HD\,15337 after removing the instrumental offset due to the change 
                     of the fibres on June 1, 2015.
                     \emph{Middle:} RV residuals after removing the best-fit model. The rms of this time-series is 1.36\,\ms. 
                     \emph{Bottom:} GLS periodogram of the RV residuals that does not highlight any significant signals.}
        \label{fig:5}      
\end{figure*}

The shape and correlation of the marginalised posterior PDFs of the parameters are shown in Fig.\,\ref{fig:app2} for the drift, offset polynomial, and GP amplitude parameter and
in Figs\,\ref{fig:app3} and\,\ref{fig:app4} for the parameters of TOI-402.01 and TOI-402.02, respectively. We observe strong correlations between the offset and the polynomial parameters, 
however the marginalised posterior PDFs are well sampled. We note that those correlations appear even though we re-centred the time-series around the time of transit\footnote{We removed from the times of 
observation the value BJD$_{UTC}$ 2458412} found by \emph{TESS} before sampling the posterior PDF. For all the other parameters
no strong correlation is observed and the marginalised posterior PDFs are also well sampled. The mean of the marginalised posterior PDF of each parameter is reported in Table\,\ref{tab:1} with each error being the rms of each marginalised posterior PDF, therefore 
describing the 68\% confidence level of each parameter. Our RV model using those parameters is shown in the top panel of Fig.\,\ref{fig:5} over-plotted on the data. In the middle  and bottom panels of the same figure, 
we can see the RV residuals and their corresponding GLS periodogram, respectively. As we can see, no significant signal appears in those RV residuals. The strongest signal is at 6.6 days, with a
$p$-value of 3.4\% and an estimated amplitude of 0.44\,\ms. The rms of the RV residuals is 1.36\,\ms, much lower than the 2.17\,\ms obtained with our preliminary RV analysis in Sect.\,\ref{sec:preliminary_analysis}. Using
a GP regression with a quasi-periodic kernel therefore much more effectively mitigates stellar activity than using a linear dependence with the high-frequency \logrhk\,time-series.

\begin{figure*}[h]
        \centering
        \includegraphics[width=18cm]{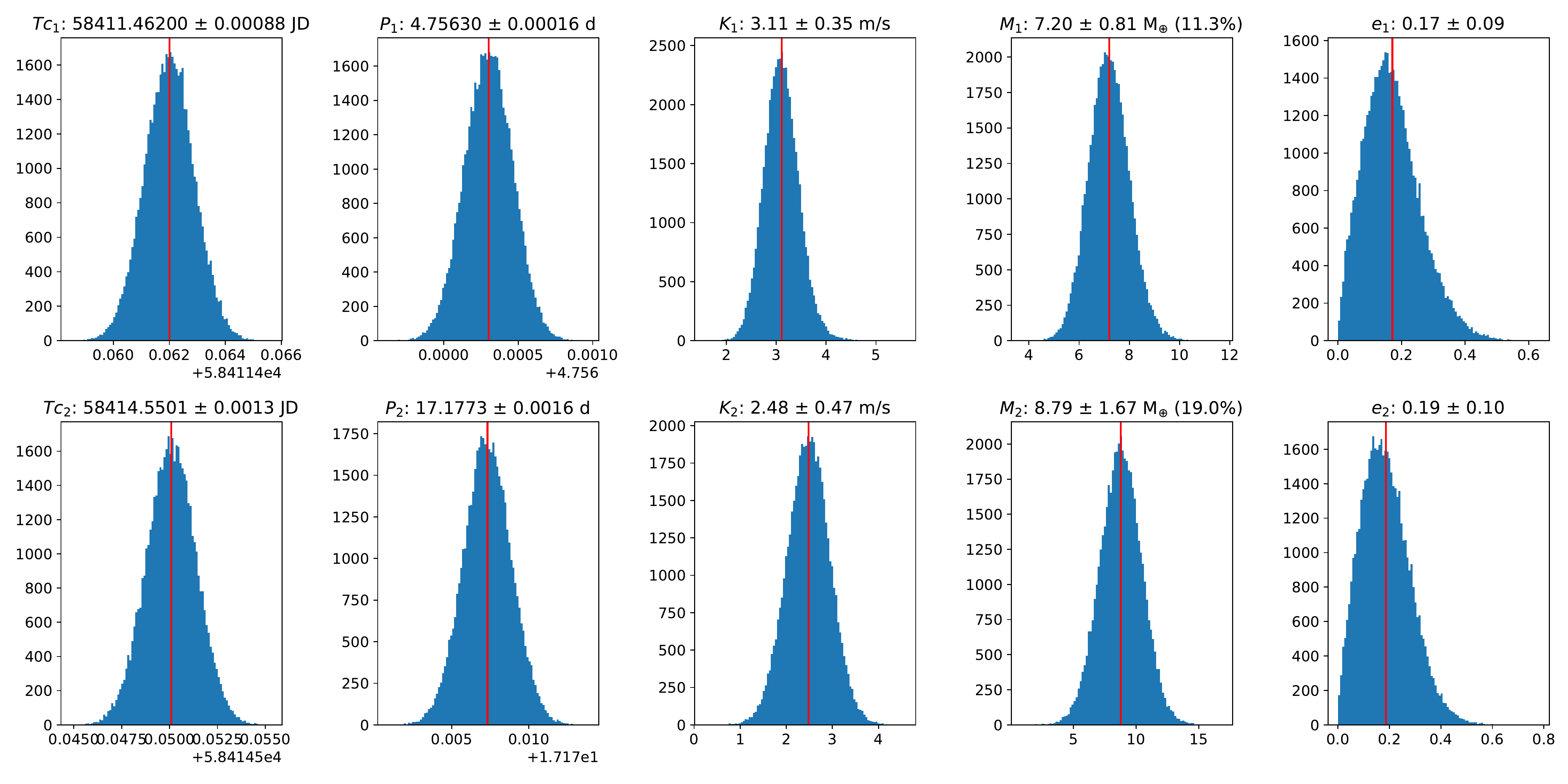}
        \caption{Marginalised posterior PDFs of the orbital parameters of TOI-402.01 (subscript 1) and TOI-402.02 (subscript 2). The planetary mass posterior PDFs have been derived by considering the orbital inclinations of TOI-402.01 and TOI-402.02 with their respective errors (see Table\,\ref{tab:phot_par}), and the stellar mass with its respective error (see Table\,\ref{tab:0}). The red vertical lines represent the mean value of each distribution. 
                      The title of each subplot gives this mean value for each marginalised posterior PDF, in addition to the rms of the distribution giving therefore the 68\% confidence interval error.}
        \label{fig:6}      
\end{figure*}

\begin{figure*}[h]
        \centering
        \includegraphics[width=9cm]{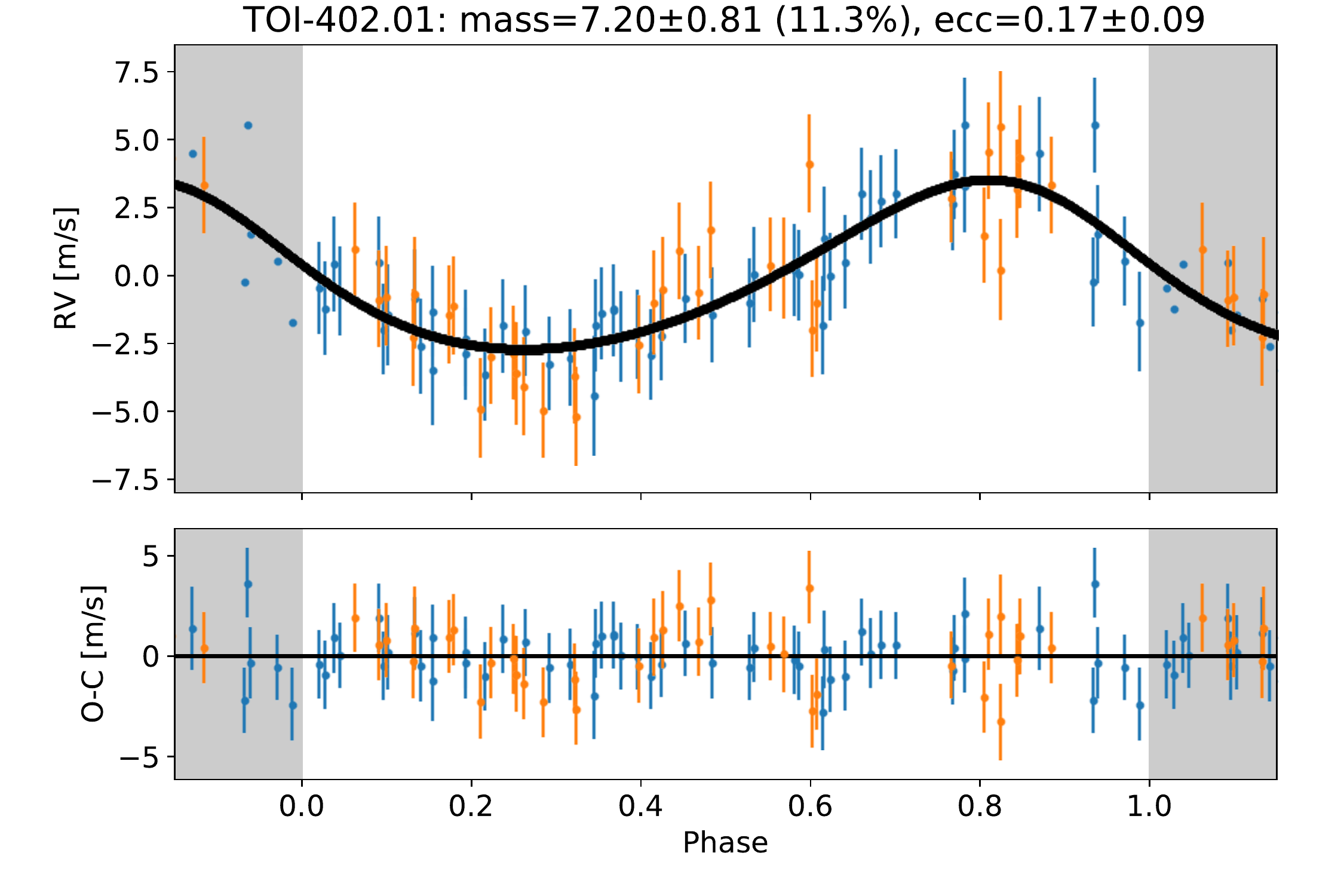}
        \includegraphics[width=9cm]{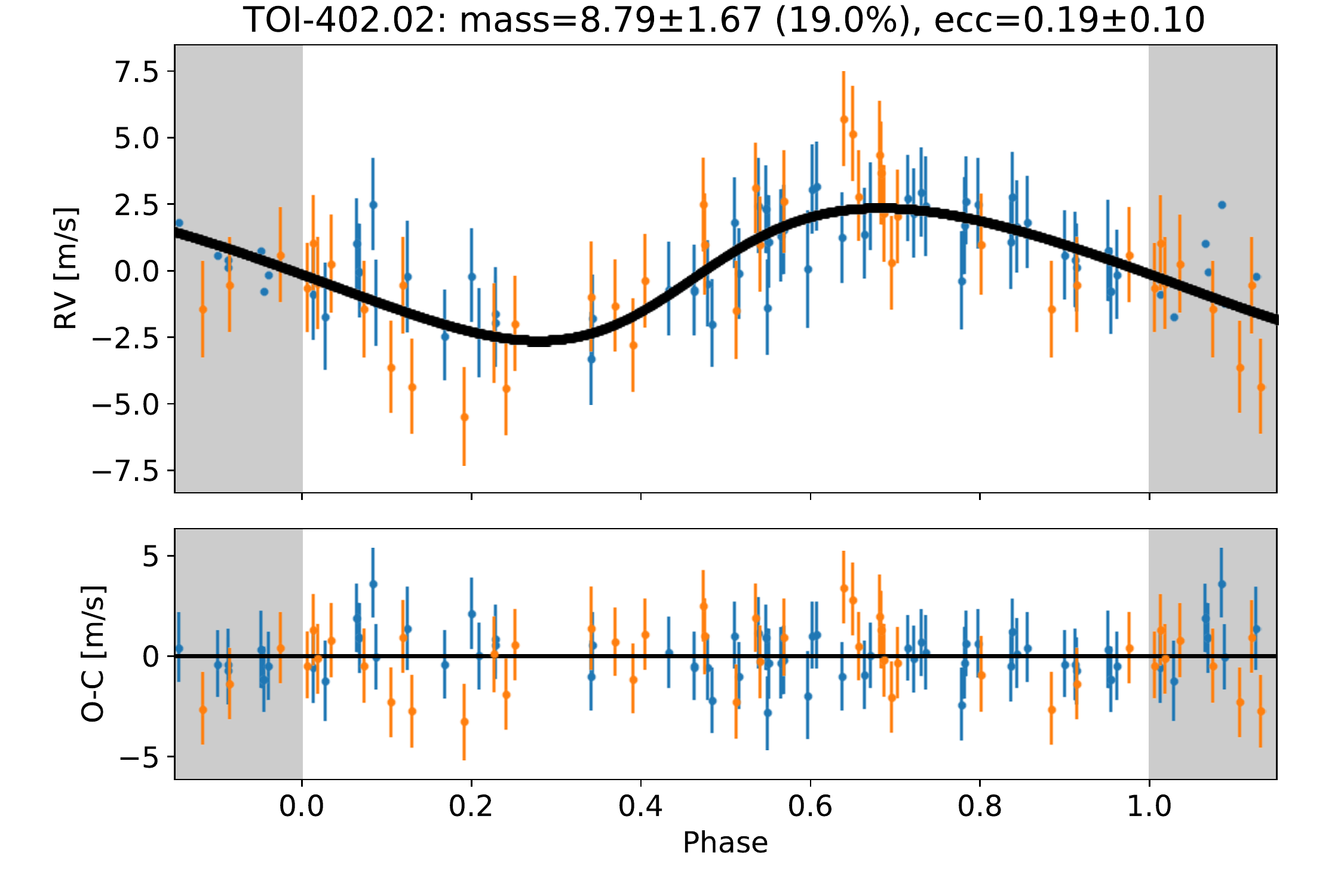}
        \caption{Phase-curves for TOI-402.01 and TOI-402. The top plots show the best fitted planetary signal in the RV data after removing the instrument offsets, the second-order polynomial drift, and the GP regression. 
                      The bottom plots show the RV residuals after removing the planetary signal.}
        \label{fig:7}      
\end{figure*}

The marginalised posterior PDFs of the planetary parameters of interest are shown in Fig.\,\ref{fig:6}. All the marginalised posterior PDFs present a Gaussian shape, except for the eccentricity that shows a slight but relatively marginal skewness, and therefore we can confidently select the mean and standard deviation as the best estimate and 68\% confident error bar for each planet parameter. Using those means, the planetary signals for TOI-402.01 and 402.02 are shown folded in 
phase in Fig.\,\ref{fig:7}. The marginalised posterior PDFs for the time of transit and period of both planets are extremely similar to the priors used, as expected, implying that the RV measurements do not constrain  those parameters
any further. All the other parameters are compatible within 1$\sigma$ with the preliminary RV analysis performed in Sect.\,\ref{sec:preliminary_analysis} and shown in Fig.\,\ref{fig:3}. Therefore, even if we consider in this section 
stellar activity with a better model, which is confirmed by the smaller rms of the RV residuals that we obtain, that is, 1.36 compared to 2.17\,\ms, the impact on the derived planetary parameters is not significant.

\begin{figure}[h]
        \centering
        \includegraphics[width=9cm]{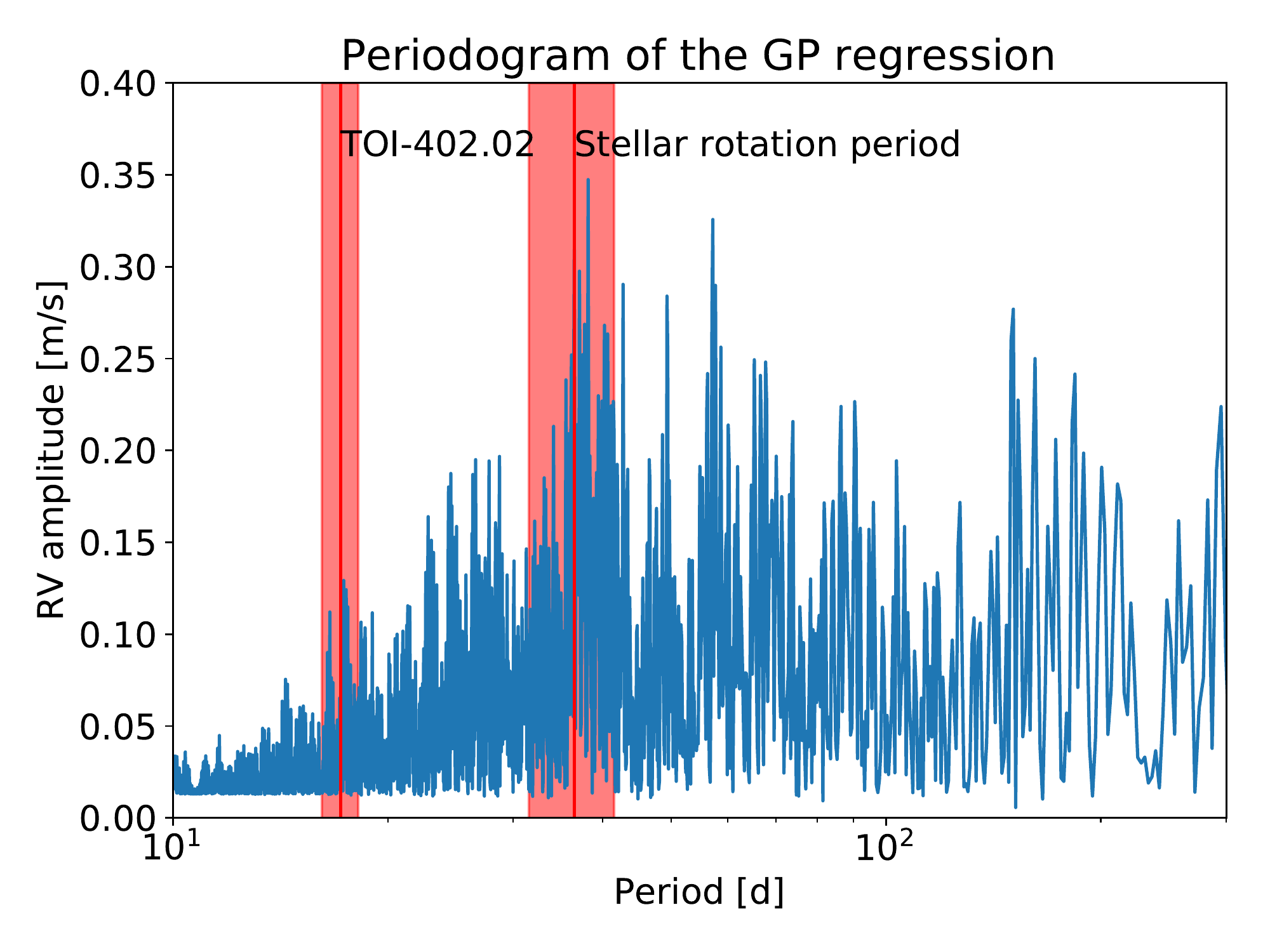}
        \caption{GLS periodogram of the GP regression showing the RV amplitude in meters per second with respect to the period. The red vertical windows highlight the orbital period of TOI-402.02 and the stellar rotation period.}
        \label{fig:8}      
\end{figure}

To verify that the GP regression fitted here does not significantly absorb the signal of TOI-402.02, we show in Fig.\,\ref{fig:8} its corresponding periodogram. As expected, the highest signals modelled by the GP regression are found at 
the stellar rotation period and at longer periods. At 17.17 days, corresponding to the orbital period of TOI-402.02, the GP regression absorbs about 0.13\,\ms. The impact of the GP on the planetary signal of TOI-402.02 is therefore negligible and we can be confident in the orbital parameters derived for this planet.

We note that we also tried to adjust the same model as presented here, this time using a GP regression with a QP kernel whose hyper-parameters were fixed to the values described in the \logrhk\,analysis (see Sect.~\ref{sec:GP_training_Rhk} and top of Table~\ref{tab:1}). However, like in the \logrhk\,analysis, we obtained values extremely similar to the fit using a SE kernel, well within 1$\sigma$. Therefore, we decided in this section to only show the results of the SE kernel as it is a simpler model.

\subsection{Combined photometric and RV analysis}
\label{sec:RV_analysis}

To test if a combined photometric and RV analysis could better constrain the planetary orbital parameters, we performed a combined modelling using EXOFASTv2. Unfortunately, as GP regression is not implemented in EXOFASTv2, this analysis does not model stellar activity.

This combined analysis slightly improves the precision on the orbital period of TOI-402.01 and TOI-402.02 by 16 and 7\%, respectively, on the time of transit of TOI-402.01 by 1\%, and on the radius of TOI-402.01 by 7\%. Also, the eccentricity of both planets slightly decreases to 0.09 instead of 0.17 and 0.19 for TOI-402.01 and TOI-402.02, respectively. However, regarding the planetary masses, the precision becomes worse by 26 and 33\%, respectively, due to the fact that stellar activity is not accounted for in this combined analysis. All the values for the different fitted parameters in this combined analysis are well within 1$\sigma$ of the values found in our analysis of the RVs using as priors the tight constraints given by photometry (see Sect.\,\ref{sec:GP_fit_RVs} and results in Table\,\ref{tab:1}).

From this analysis, we can conclude that the benefit of performing a combined photometric and RV analysis including GP regression would be marginal. We are therefore confident that the mass, radius, and orbital characteristics of TOI-402.01 and TOI-402.02 stated in this paper are robust.

\section{Discussion}
\label{sec:discussion}

\subsection{Interior characterisation}
\label{sec:interior}

Given the bulk densities, TOI-402.01 is dominated by a rocky composition, while TOI-402.02 must contain volatile layers. We performed an analysis similar to \citet{dorn2017generalized} to further constrain the planet bulk properties. All the details of the analysis and key figures can be found in Appendix\,\ref{app:interior}.
Figure \ref{plot_interior} and Table \ref{tableresults} summarise posterior distributions of inferred interior parameters for both TOI-402.01 and TOI-402.02. The bulk densities of the two planets are significantly different. In consequence, the interiors of  TOI-402.01 have negligible amounts of volatile elements (i.e. water and gas), while there  are significant amounts for TOI-402.02.

TOI-402.01 is predominantly rocky (\rsolid = $0.99_{-0.06}^{+0.01}$ R$_p$). The mass, radius, and bulk abundances provide information on core size and mantle composition. We note that the bulk abundance constraints do not allow the interiors to match the measured bulk density of $\rho_p=1.72 \rho_\oplus$. This is because the abundance constraints favour Earth-like densities, while the bulk density of 
TOI-402.01 is higher. In order to better fit the bulk density, we relaxed the constraint on Fe/Si in a separate scenario and thereby allowed for rocky interiors with large core mass fractions (Table \ref{tableresults}). Although this scenario can fit mass and radius, it remains unclear how such iron-rich interiors for massive super-Earths can be formed. We note that the addition of light elements in the core might be crucial to reconcile measured abundance constraints with bulk density. Here, we have assumed pure iron cores for simplicity. Further investigations are required to understand the importance of light core elements for super-Earths.

TOI-402.02 hosts a thick atmosphere that is likely dominated by primordial H/He.
Although, mass and radius are insufficient to determine the composition of the gas envelopes (uncertainties on \Zenv are large), considering the evolution of the atmospheres can inform us on the nature of the gas as stellar irradiation drives atmospheric escape and can efficiently erode thin primordial atmospheres of H/He on short time-scales \citep{dorn2018secondary}. In consequence, primordial H/He gas can be excluded for atmospheres that are too thin in order to be stable against evaporative loss.
Following the work of \citet{dorn2018secondary}, there is a theoretical minimum threshold thickness for a primordial H/He atmosphere, which corresponds to the amount of gas (in H$_2$) that is lost on a short time-scale (here we use 100 Myr). For TOI-402.01, assuming solar-like X-ray and UV (XUV) flux emitted by the host star, this threshold thickness is 0.12 $R_{\rm p}$ and thus larger than the inferred thicknesses ($r_{\rm env} = 0.01_{-0.01}^{+0.05} R_{\rm p}$). Any H/He-dominated layer is therefore excluded for TOI-402.01 and a possible thin atmosphere could be of volcanic origin.
In the case of TOI-402.02, the threshold thickness is  0.07 $R_{\rm p}$ and inferred thicknesses are larger ($r_{\rm env} = 0.1_{-0.05}^{+0.06} R_{\rm p}$) which indicates that a primordial H/He layer on TOI-402.02 is likely for this less-irradiated planet.

If TOI-402.01 has indeed an atmosphere, this would be interesting from a theoretical point of view, since outgassing processes are very mass dependent for some tectonic regimes (e.g. stagnant-lid).  In the case of stagnant-lid regimes, no massive terrestrial-like atmospheres are expected for planets of $\gtrapprox$ 8 $M_\oplus$  \citep{noack2017volcanism, dorn2018outgassing}. 
A massive atmosphere of volcanic origin could only be present if the planet is in a different convection regime, for example with plate tectonics. Whether super-Earths can drive plate tectonics or not is still unclear from the variety of modelling studies \citep{valencia2007inevitability, noack2014plate, korenaga2010likelihood, van2011plate}. 

\subsection{Temporal evolution under XUV-driven atmospheric escape}
\label{sec:temp_evo}

The estimates in the previous section suggest that TOI-402.01 should have lost a primordial H/He envelope, whereas TOI-402.02 should have kept it. The same conclusion is reached when comparing the location of the planets in the semimajor axis--radius and semimajor axis--mass planes directly with the locus of the evaporation valley as predicted by evaporation models \citep{owenwu2013,lopezfortney2013,jin2014planetary}. In \citet{Jin:2018aa}, the authors predict that for solar-like star hosts, Earth-like bulk composition planets less massive than $M_{\rm core}$$\approx$$6\,(a/0.1 {\rm AU})^{-1}\,\Mearth$, which corresponds to  $R_{\rm core}$$\approx$$1.6\,(a/0.1 {\rm AU} )^{-0.27}\,\Rearth$, loose their entire H/He envelope because of XUV-driven atmospheric escape. For the low-mass close-in planets orbiting TOI-402, the total mass is approximately the same as the (astrophysical) core mass \citep{lopezfortney2014}. Therefore, using the theoretical prediction from \citet{Jin:2018aa}, we find that $M_{\rm core}$ should be at least $\sim$ 11.5 \,\Mearth\,for TOI-402.01 to keep its H/He envelope, while it should be at least $\sim$5\,\Mearth\,for TOI-402.02. With masses of 7.20 and 8.79\,\Mearth\,for TOI-402.01 and TOI-402.02, respectively, we thus arrive at the conclusion that the inner planet lost its entire H/He envelope, while its outer companion kept a significant part of it. 
Therefore, these two planets should be found on both sides of the evaporation valley, which is the case given their distinct densities, as discussed above (see also Fig. \ref{fig:atm}). 

\label{sec:temporal_evolution}
\begin{figure*}
        \centering
        \includegraphics[width=\textwidth]{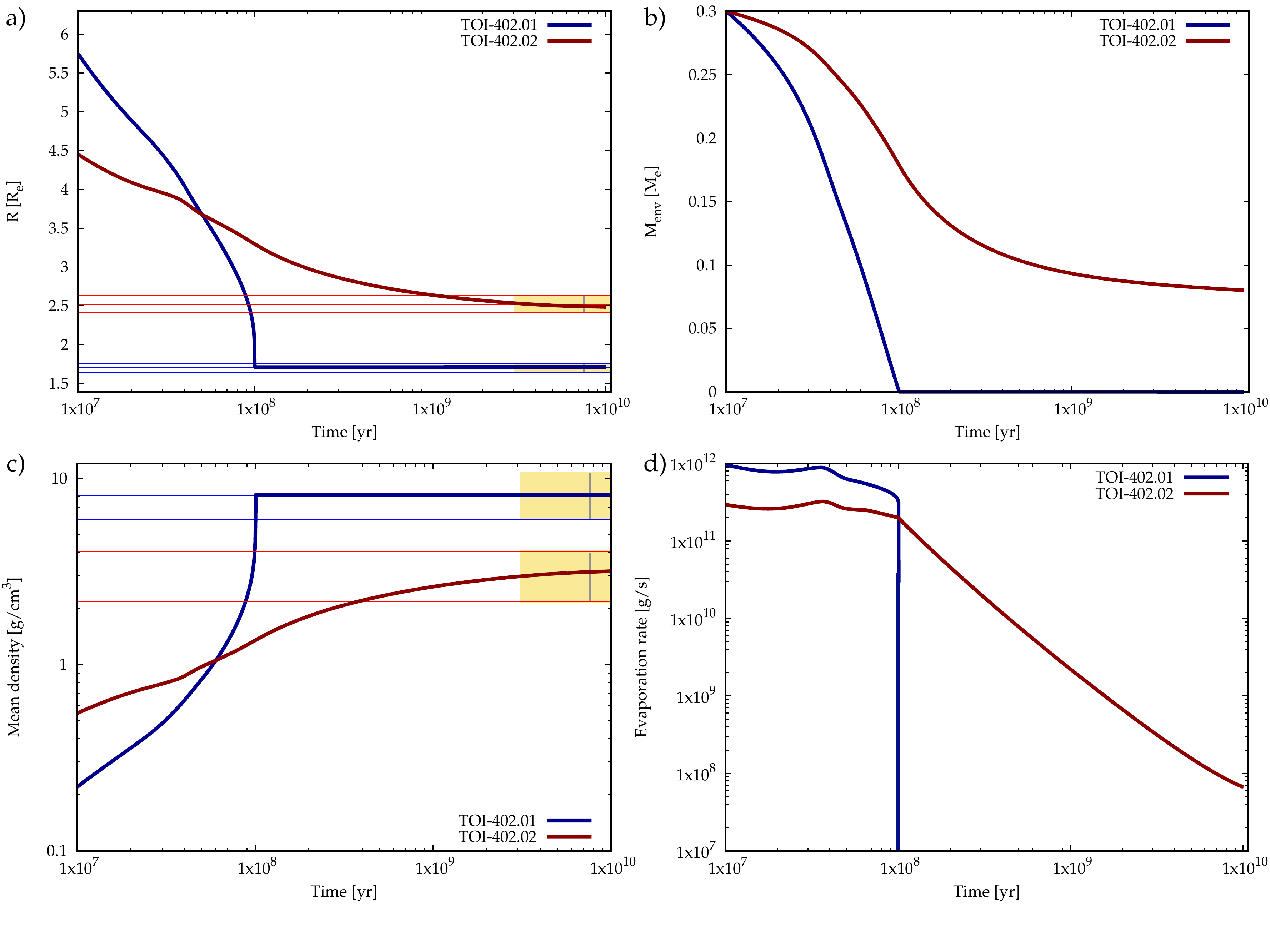}
        \caption{Temporal evolution of the TOI-402 planets. The panels show as a function of time (a) the transit radius (in Earth radii), (b) the mass of the H/He envelope (in Earth masses), (c) the mean density of the planets, and (d) the XUV-driven hydrodynamic H/He evaporation rate. In the two panels on the left, the horizontal lines show the observed  values and the 1-$\sigma$ error intervals. The yellow region indicates the  1-$\sigma$ age range of the star, with the vertical grey line showing the nominal age of 7.5 Gyr. }
        \label{fig:evo}      
\end{figure*}

To directly test the hypothesis that evaporation has shaped the distinct densities of the planets, we have  simulated the long-term thermodynamical evolution (cooling, contraction, atmospheric escape) of the two planets. The goal is to understand first whether the observed properties of the planets (like in particular the large density contrast) can indeed be explained by  evolution under XUV-driven evaporation, and second to constrain the post formation properties. 
Similar studies \citep{lopezfortney2013,owenmorton2015} were conducted in the past for the planets around Kepler-36 \citep{carteragol2012} which also have a large density contrast.

To do so, we simulated the evolution of the planets using the Bern planet evolution model \texttt{completo21} which was described in \citet{mordasinialibert2012b,jin2014planetary,lindermordasini2018}. XUV-driven escape in the radiation-recombination and energy-limited regimes is considered \citep{murray-claychiang2009}. The cores have an Earth-like 2:1 silicate:iron composition described by the polytropic EOS of \citet{seager2007}. The envelope consists of H/He described by the EOS of \citet{saumonchabrier1995}. The opacity corresponds to a condensate-free solar-composition gas \citep{freedmanlustig-yaeger2014}.

Figure \ref{fig:evo} shows a simulation that reproduces the observed masses and radii at the observed age of the star. The initial conditions, the final results, and the comparisons to the observed values are given in Table \ref{tab:evo}. The most important outcome, that is, that TOI-402.01 becomes a bare core and TOI-402.02 keeps H/He, is immediately found without tuning  the initial conditions or model parameters away from generic values (i.e. an initial envelope mass of 1 to 10\% of the core mass, an efficiency factor of atmospheric escape of about 0.1 to 0.3, and a saturated XUV luminosity of about $10^{-3}$ of the bolometric luminosity).
\begin{table*}[ht]
        \scriptsize
        \caption{Initial conditions and results of the evolutionary calculations, and comparison to the observed values.}            
        \label{tab:evo}     
        \centering                         
        \begin{tabular}{l ccc|ccc}       
                \hline\hline
                Planet & \multicolumn{3}{c}{\bf{TOI-402.01}} &  \multicolumn{3}{c}{\bf{TOI-402.02}}\\
                Quantity & Post-formation & 7.5 Gyr & Observation        & Post-formation & 7.5 Gyr & Observation         \\
                Total mass [$\Mearth$] & 7.50 & 7.20 &  7.20$\pm$0.81& 9.03 & 8.81 & 8.79$\pm$1.68\\
              H/He envelope mass [$\Mearth$]  & 0.30 & 0.00  & - & 0.30 & 0.08 & - \\
              Total radius [$\Rearth$] & 5.73 & 1.71 & 1.70$\pm$0.06 & 4.45 & 2.49 & 2.52$\pm$0.11 \\
              Mean density [g/cm$^{3}$] & 0.22 & 8.16 & 8.05$^{+1.93}_{-1.61}$ & 0.55 & 3.14 & 3.02$^{+1.09}_{-0.87}$\\        
              H/He evaporation rate [g/s] & 1$\times$10$^{12}$& 0.00 & - & 3$\times$10$^{11}$ & 9$\times$10$^{7}$& -             
\end{tabular}
\end{table*}

The only initial condition that requires some fine tuning is the initial (i.e. post-formation) envelope mass of TOI-402.02. The value of 0.3 $\Mearth$ was determined by requiring that the transit radius at 7.5 Gyr correspond to the observed value. This value is on the lower side of  what planet-formation models find for a planet with a (core) mass of about 9 $\Mearth$ \citep[e.g.][]{alibertmordasini2005,ikomahori2012,mordasini2018}, but still well within the range predicted by models explicitly solving the governing internal structure equations (see \citealt{mordasini2018}). 

For TOI-402.01, the post-formation H/He envelope mass is in contrast not important for the final outcome (one always finds a complete loss of the envelope) and thus remains unconstrained. In the figure, the same value as for TOI-402.02, 0.3 $\Mearth$, was used for simplicity. Choosing for example a much higher initial mass of 1.6 $\Mearth$ (with the same core mass) also leads to a loss of the envelope, but at a later time of about 400 Myr. The reason for this independency is that a higher initial envelope mass means that on one hand, there is more mass to evaporate, but on the other hand, the mean density of the planet is also lower \citep{lopezfortney2014}, which in turn means that the evaporation rate is higher.
This drives the system to an outcome that is independent of the post-formation envelope mass.

In the top-left panel of Fig. \ref{fig:evo}, the radii are initially significantly larger than today. This is mainly due to the high intrinsic luminosity of young planets \citep{mordasinimarleau2017}.  The radius of TOI-402.01  is initially larger than that of TOI-402.02, which is a consequence of the identical initial envelope masses and the  lower (core) mass of  TOI-402.01. The rapid decrease of the radius of TOI-402.01 at the moment when the last bit of H/He is lost (at about 100 Myr) is also visible. This significant decrease of the radius on a short timescale gives rise to the existence of the evaporation valley  on a population-wide level \citep{jin2014planetary}. The bottom-left panel shows the corresponding mean densities.

The top-right panel displays the decrease of the H/He envelope mass. We see that the envelope mass of TOI-402.02 today is about 0.08 $\Mearth$, about a third of the initial mass. Most of the escape occurs as expected early on \citep[e.g.][]{lopezfortney2013}, when the stellar XUV luminosity is high, and the planetary radii are large.

The bottom-right panel finally shows the total H/He evaporation rate as a function of time for the two planets. During the saturated phase - assumed to have a duration of 100 Myr \citep[e.g.,][]{tujohnstone2015} - the  evaporation rates are high, about $3 \times 10^{11}$\,g\,s$^{-1}$ for the outer planet, and even almost $10^{12}$\,g\,s$^{-1}$ for the inner one. The evaporation rate of TOI-402.02 nowadays is predicted to be close to $10^{8}$\,g\,s$^{-1}$. For comparison, the escape rate of the warm Neptune-mass planet GJ 3470b was recently measured to be about $10^{10}$\,g\,s$^{-1}$ \citep{bourrierlecavelier2018}. 

In summary, we conclude that the current-day densities of the planets and in particular the large density contrast between them can be  well explained by atmospheric escape of H/He envelopes surrounding  Earth-like cores.  The same conclusion was already reached for another important system with planets on both side of the valley, Kepler-36 \citep{lopezfortney2013,owenmorton2015}. There is however one interesting difference between Kepler-36 and TOI-402: in Kepler-36, the two planets are at very similar orbital distances (difference of just 11\%), but differ clearly in mass (difference of about 80\%). This mass difference is the reason that the outer planet around Kepler-36 can keep the H/He. 

Here, in TOI-402, the mass difference is in contrast much smaller (only about 20\%), but the outer planet resides at a much larger orbital distance (difference of 240\%). Here, it is because of this larger orbital distance combined with the negative radial slope of the evaporation valley \citep{vaneylen2018} that the outer planet can keep its H/He. The  somewhat higher mass of TOI-402.02  alone would in contrast not suffice to prevent complete loss if the planet were to be placed at (or close to) the  position of TOI-402.01.


\subsection{Atmospheric characterisation}
\label{sec:atmos_characterisation}

Although the present atmosphere of TOI-402.02 strongly depends on XUV irradiation history (\citep[][]{vidal-madjar_extended_2003}), our analysis of its past evolution under stellar irradiation shows that it could have survived atmospheric escape and kept a substantial fraction of its primordial volatile envelope, about a third of its total mass (see Fig.~\ref{fig:evo} and Sect.~\ref{sec:temp_evo}). Atmospheric mass loss from TOI-402.02 decreased over time as the host star became less luminous, the planet density increased (making it more difficult for gas to escape), and the radius of its envelope decreased (decreasing the area over which it captures stellar energy). Nonetheless, its atmosphere could  still escape today at a moderate rate of about 10$^{8}$\,g\,s$^{-1}$.

Whether TOI-402.02 is water-rich or hosts a H/He envelope, the composition of its escaping outflow will be dominated by hydrogen atoms. The planet is sufficiently irradiated for water to turn into steam, which would photo-dissociate at high altitudes and sustain a hydrogen-rich upper atmosphere prone to escape (\citep[][]{wu_velocity_1993, jura_observational_2004}). The resulting exosphere could be searched for using transit spectroscopy in the ultraviolet stellar Lyman-$\alpha$ line, which is sensitive to the neutral hydrogen atoms. The neutral fraction of the outflow depends on several processes and properties (photoionization, temperature, planetary density,  etc.), but a super-Earth could have upper atmospheres that are almost completely neutral \citep[][]{Salz2016b}. Escape rates of neutral hydrogen on the order of 10$^{8}$\,g\,s$^{-1}$ from TOI-402.02 could be readily detected in a few transits with the Hubble Space Telescope (see the case of GJ\,436b, \citealt{Bourrier2016}; and HD\,97658b, \citealt{bourrier_no_2017}, a planet with similar properties and host star to TOI-402.02; see Fig.~\ref{fig:atm}). The main limitation in observing stellar Lyman-$\alpha$ lines resides in their absorption by the interstellar medium. The proximity of TOI-402 (45\,pc), combined with its brightness and its stellar type, makes it one of the few stars to host small transiting planets that would allow for Lyman-$\alpha$  transit spectroscopy to be carried out using the Hubble Space Telescope.

The expected signal of water vapour in transmission, for five scale heights and assuming a bond albedo of zero and a mean molecular weight of 20, is around a few parts per million for both TOI-402 planets. This signal is out of reach of present instrumentation, but could be detectable in the near future with the James Webb Space Telescope or future ground-based Extremely Large Telescopes. Searching for the signature of helium in the upper atmosphere of TOI-402.02 could however be done with current facilities to disentangle between a water-dominated and a H/He envelope. Helium in the escaping outflow could be detected via its near-infrared triplet, a new atmospheric tracer \citep[][]{Seager:2000aa, Oklopcic:2018aa} that can yield deep absorption signatures when observed at high spectral resolution \citep[][]{allart_spectrally_2018, nortmann_ground-based_2018, salz_detection_2018, allart_high-resolution_2019}. 
\section{Conclusion}
\label{sec:conclusion}

We confirm the detection of the two planets transiting TOI-402 found by \emph{TESS} using HARPS RV measurements. With periods of 4.75642$\pm$0.00021 and 17.1784$\pm$0.0016 days,
radii of 1.70$\pm$0.06 and 2.52$\pm$0.11\,\Rearth\,, and masses of 7.20$\pm$0.81 and 8.79$\pm$1.68 \Mearth, we find that one planet is a hot rocky super-Earth 
and the other a warm puffy super-Earth. The precision on the planetary radii is 3.6 and 4.2\%, respectively. The precision on these planetary masses is 11.3 and 19.1\%, including the error in the host stellar mass of 4.9\% and in the orbital inclination of 0.4 and 0.06\%, respectively.
Those values have been derived by re-analysing the \emph{TESS} photometry using EXOFASTv2, and analysing 85 HARPS 
public RV measurements. The photometric and spectroscopic analyses were performed separately, however the tight marginalised posterior PDFs for the period and
time of transit of TOI-402.01 and TOI-402.02 obtained from photometry were used as priors to the RV analysis. We note that the host star TOI-402 (HD\,15337), a K1 dwarf with effective temperature 5131$\pm$74\,K and $\log{g}$ 4.37$\pm$0.13,
is a rather low-activity star like the Sun and its RVs are affected by stellar signals at the 2\,\ms level (see the amplitude of the GP regression in Table\,\ref{tab:1}). We therefore used a GP regression to model stellar activity in the RV time-series.

\begin{figure}[h]
        \centering
        \includegraphics[width=9cm]{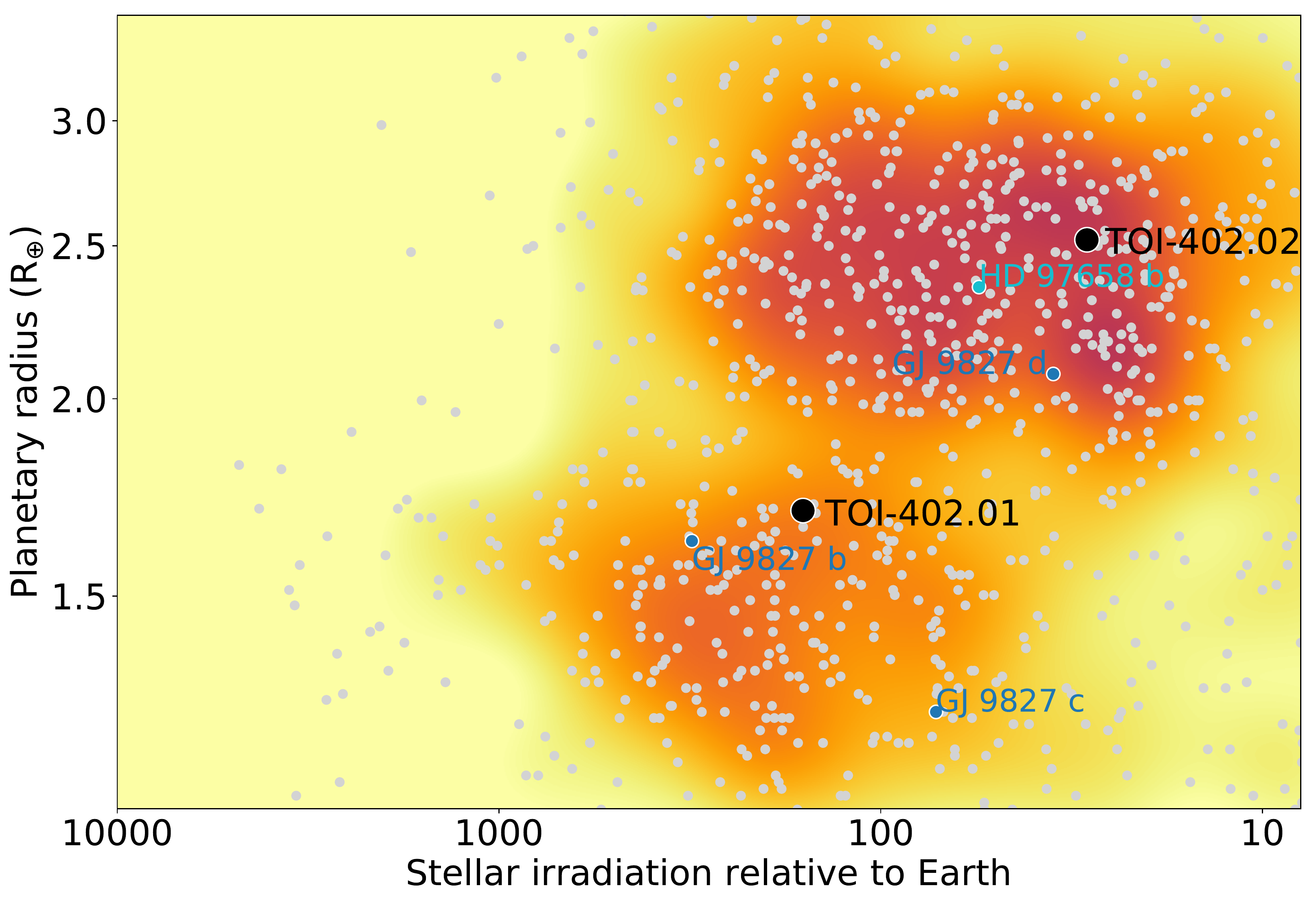}
        \caption{Planetary radius as a function of the stellar irradiation received by the planet \citep[based on][]{Fulton:2017aa}. TOI-402.01 stands just below the radius gap while TOI-402.02 stands above. The GJ\,9827 system is added here as it represents, with TOI-402, probably the best system to understand the radius gap through exoplanetology comparison. HD\,97658b is also added for a direct comparison with TOI-402.02 as they are supposed to have water-rich composition and, furthermore, a search for photo-dissociation of water has been undertaken on HD\,97658b by \citep[][]{bourrier_no_2017}. The data at the origin of this diagram have been extracted from the NASA exoplanet archive (\url{https://exoplanetarchive.ipac.caltech.edu}).}
        \label{fig:atm}      
\end{figure}

As we can see in the irradiation-radius diagram shown in Fig.\,\ref{fig:atm}, the two planets lie on different sides of the radius gap found at $\sim$1.8\,\Rearth\,\citep[][]{Buchhave-2014, Rogers-2015, Fulton:2017aa}. The origin of this gap is not yet clearly understood \citep[e.g.][]{Lehmer:2017aa,Ginzburg:2018aa}. Photo evaporation is likely the main driver \citep[][]{jin2014planetary,Owen:2017aa,Jin:2018aa}, and the study of the temporal evolution under XUV-driven atmospheric escape of the TOI-402 planetary system performed in this paper confirms this hypothesis (see Sect.~\ref{sec:temp_evo}).

With the brightness of the host star (V$=9.09$ and J$=7.55$), TOI-402 is one of the most amenable systems, with GJ\,9827 \citep[][]{Niraula:2017aa, rodriguez_system_2018}, to perform comparative exoplanetology across the evaporation valley and thus help us to better understand the radius gap.

\begin{acknowledgements}
XD is grateful to The Branco Weiss Fellowship--Society in Science for its financial support. 
This study was based on observations collected at the European Southern Observatory under the GTO program 072.C-0488(E) and Large programs 183.C-0972(A), 192.C-0852(A), 196.C-1006 and 198.C-0836(A).
We acknowledge the use of public TESS Alert data from pipelines at the TESS Science Office and at the TESS Science Processing Operations Center.
Resources supporting this work were provided by the NASA High-End Computing (HEC) Program through the NASA Advanced Supercomputing (NAS) Division at Ames Research Center for the production of the SPOC data products.
We thank the Swiss National Science Foundation (SNSF) and the Geneva University for their continuous support to our planet search programs. This work has been in particular carried out in the framework of the \emph{PlanetS} National Centre for Competence in Research (NCCR) supported by the SNSF. This publication makes use of the Data \& Analysis Center for Exoplanets (DACE), which is a facility based at the University of Geneva (CH) dedicated to extrasolar planets data visualisation, exchange and analysis. DACE is a platform of the PlanetS NCCR, federating the Swiss expertise in Exoplanet research. The DACE platform is available at \url{https://dace.unige.ch}. 
NSC, VA, and S.G.S were supported by FCT - Funda\c c\~ao para a Ci\^encia e a Tecnologia through national funds and by FEDER through COMPETE2020 - Programa Operacional Competitividade e Internacionaliza\c c\~ao by these grants: UID/FIS/04434/2013 \& POCI-01-0145-FEDER-007672; PTDC/FIS-AST/28953/2017 \& POCI-01-0145-FEDER-028953 and PTDC/FIS-AST/32113/2017 \& POCI-01-0145-FEDER-032113. S.G.S acknowledge support from FCT through Investigador FCT contract nr. , IF-00028-2014-CP1215-CT0002. V.A. acknowledges support from FCT through Investigador FCT contract IF/00650/2015/CP1273/CT0001. C.M. acknowledges the support from the Swiss National Science Foundation under grant BSSGI0$\_$155816 ``PlanetsInTime''.
This work has made use of data from the European Space Agency (ESA) mission Gaia \url{https://www.cosmos.esa.int/gaia}, processed by the Gaia Data Processing and Analysis Consortium (DPAC, \url{https://www. cosmos.esa.int/web/gaia/dpac/consortium}. Funding for the DPAC has been provided by national institutions, in particular the institutions participating in the Gaia Multilateral Agreement. 
This research has made use of the SIMBAD database operated at CDS, France.
\end{acknowledgements}

\bibliographystyle{aa}
\bibliography{dumusque_bibliography,mordasini_bibliography}


\begin{appendix}

\section{Correlation plots between the marginalised posterior PDFs for our modelling of the high-frequency \logrhk\,time-series}

\begin{figure*}[h]
        \centering
        \includegraphics[width=16cm]{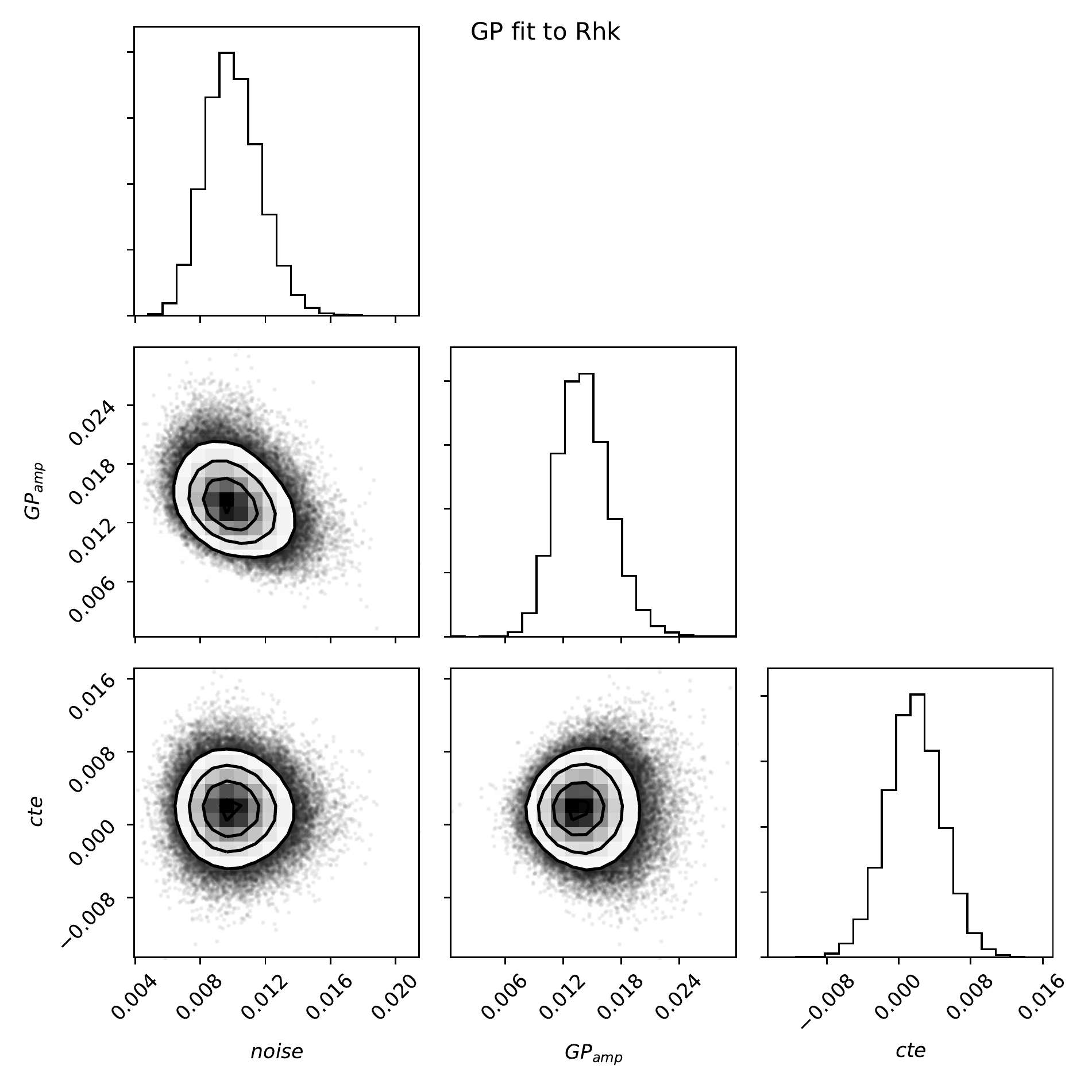}
        \caption{Correlation between the different marginalised posterior PDFs of the parameters in our modelling of the high-frequency \logrhk\,time-series.}
        \label{fig:app1}      
\end{figure*}

\clearpage
\section{Correlation plots between the marginalised posterior PDFs for our modelling of the RV time-series}

\begin{figure*}[h]
        \centering
        \includegraphics[width=16cm]{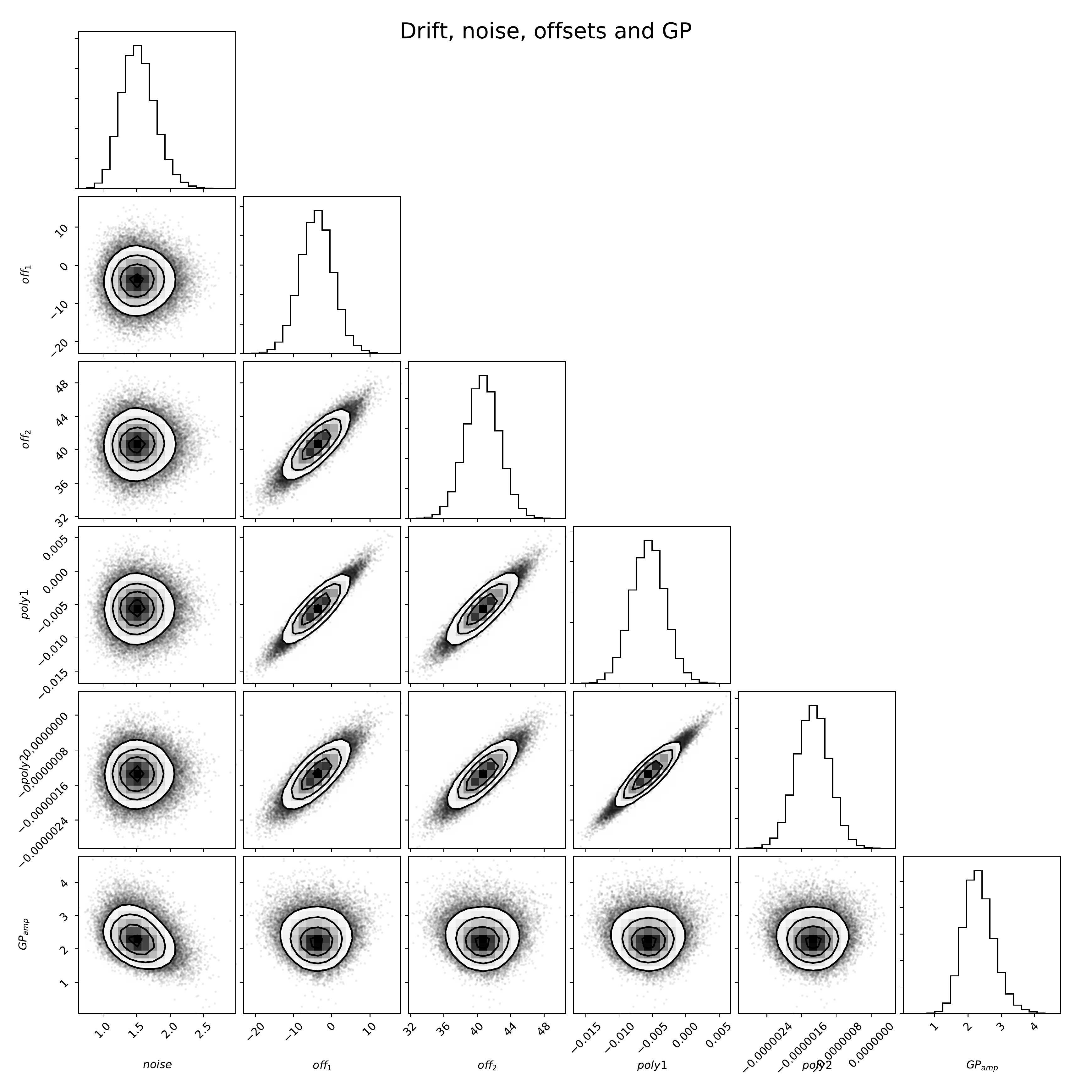}
        \caption{Correlation between the different marginalised posterior PDFs of the non-planetary parameters in our modelling of the RV time-series.}
        \label{fig:app2}      
\end{figure*}

\begin{figure*}[h]
        \centering
        \includegraphics[width=16cm]{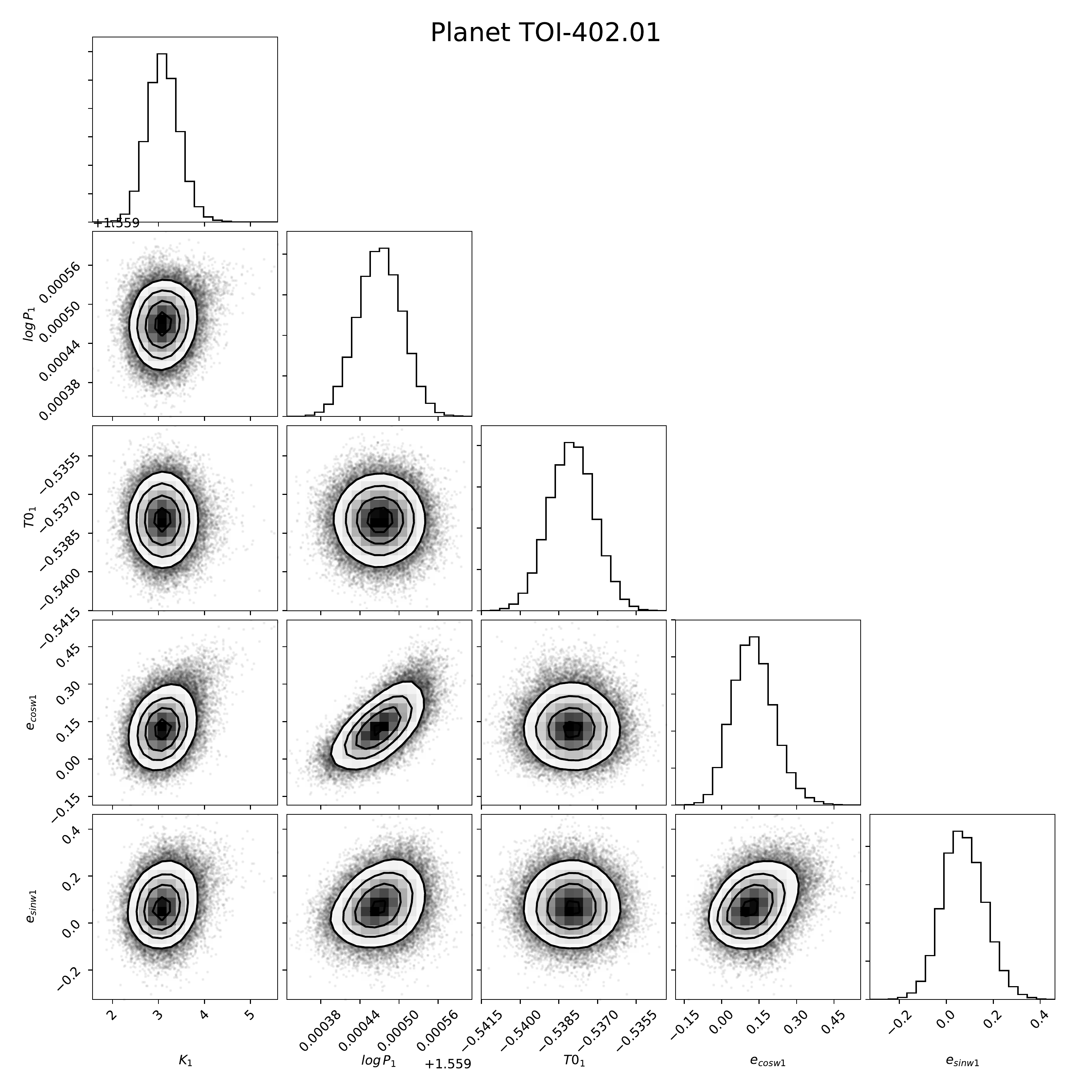}
        \caption{Correlation between the different marginalised posterior PDFs of the parameters for planet TOI-402.01.}
        \label{fig:app3}      
\end{figure*}

\begin{figure*}[h]
        \centering
        \includegraphics[width=16cm]{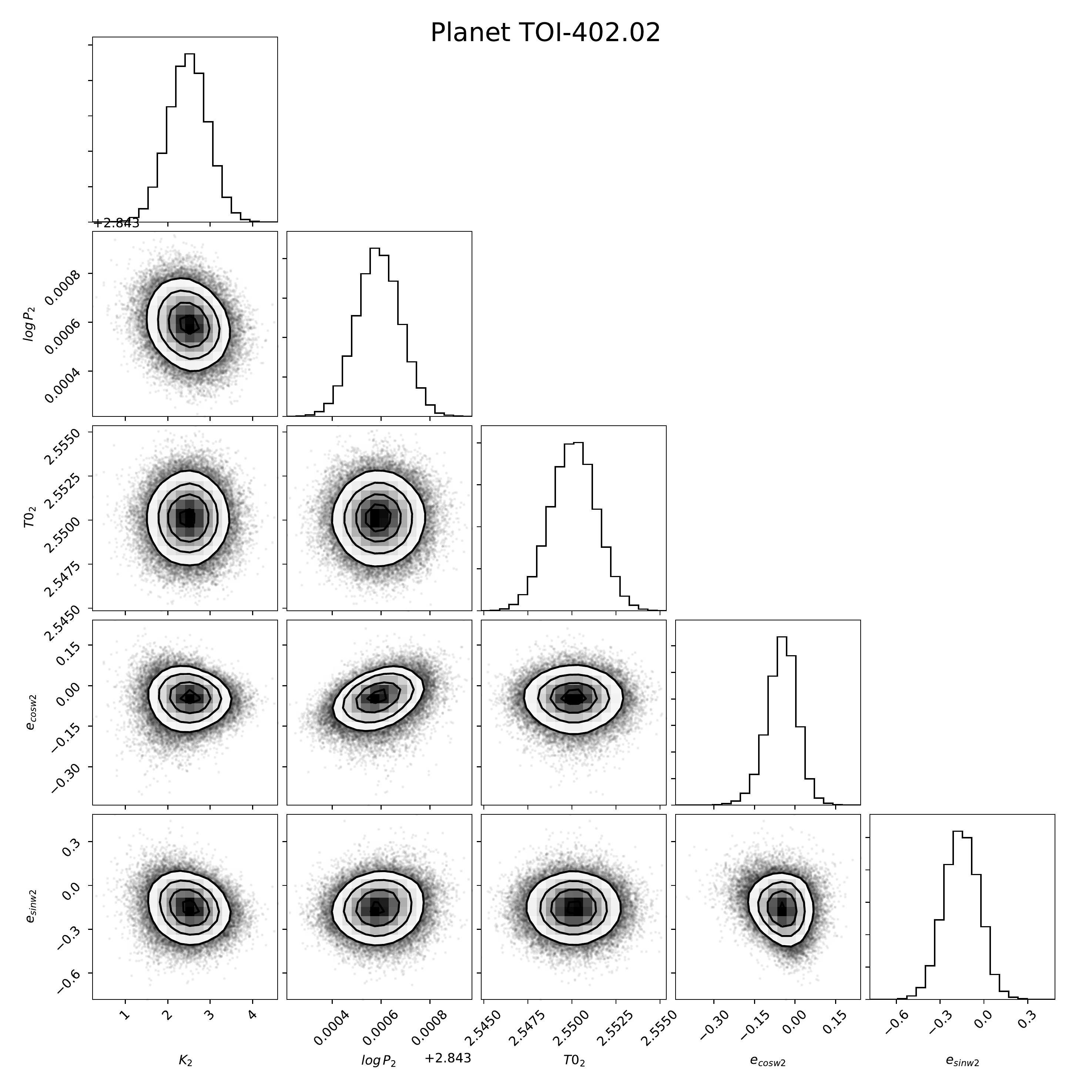}
        \caption{Correlation between the different marginalised posterior PDFs of the parameters for planet TOI-402.02.}
        \label{fig:app4}      
\end{figure*}

\clearpage
\section{Method for interior characterization}
\label{app:interior}

We use the probabilistic analysis of \citet{dorn2017generalized} for a more detailed interior characterization. 
The planetary and stellar parameters inferred in this paper are used as input to the our analysis.
Our data comprise:
\begin{itemize}
\item planet radii and masses (Tables \ref{tab:phot_par} and \ref{tab:1}),
\item planet effective temperature (Table \ref{tab:phot_par}),
\item relative stellar abundances of Fe, Si, Mg (Table \ref{tab:0}).
\end{itemize}
We note that stellar abundances are used as proxies for the planetary bulk composition \citep{dorn2015can}.

Our assumptions for the interior model are similar to those in \citet[][model I]{dorn2017generalized} and are summarised in the following. We consider planets being made of iron-rich cores, silicate mantles, layers of water ice and oceans, and an atmosphere. 
The interior parameters comprise:
\begin{itemize}
\item core size $r_{\rm core}$,
\item size of rocky interior $r_{\rm core+mantle}$,
\item mantle composition (i.e., Fe/Si$_{\rm mantle}$, Mg/Si$_{\rm mantle}$),
\item mass of water $m_{\rm water}$,
\item amount of gas $m_{\rm atm}$,
\item planet intrinsic luminosity \Lenv,
\item  gas metallicity \Zenv.
\end{itemize}
The prior distributions of the interior parameters are stated in Table \ref{tab:priorinterior}.

\begin{table*}
\caption{Prior ranges for interior parameters. $m_{\rm atm, max}$ refers to the maximum gas mass fraction based on the scaling law of \citet[equation 18 in][]{ginzburg2016super}.
 \label{tab:priorinterior}}
\begin{center}
\begin{tabular}{lll}
\hline
\hline\noalign{\smallskip}
Parameter & Prior range & Distribution  \\
\noalign{\smallskip}
\hline
\hline\noalign{\smallskip}
Core radius $r_{\rm core}$         & (0.01  -- 1) $r_{\rm core+mantle}$ &uniform in $r_{\rm core}^3$\\
Fe/Si$_{\rm mantle}$        & 0 -- Fe/Si$_{\rm star}$&uniform\\
Mg/Si$_{\rm mantle}$      & Mg/Si$_{\rm star}$ &Gaussian\\
Size of rocky interior $r_{\rm core+mantle}$   & (0.01 -- 1) $R_p$& Uniform in $r_{\rm core+mantle}^3$\\
Water mass fraction $m_{\rm water}$ & $10^{-6}$ -- 0.5 $M_p$& uniform in log-scale\\
gas mass $M_{\rm atm}$           & $10^{-11}$  -- $m_{\rm atm, max}$  &uniform in log-scale\\
planet intrinsic luminosity \Lenv            & $10^{18} - 10^{23}$ erg/s & uniform in log-scale \\
gas envelope metallicity \Zenv             & 0--1& uniform\\
\hline
\end{tabular} 
\end{center}
\end{table*}

Our interior model uses a self-consistent thermodynamic model from \citet[model I in][]{dorn2017generalized}. For any given set of interior parameters, it allows us to calculate the respective mass, radius, and bulk abundances and compare them to the actual observed data. The thermodynamic model comprises the equation of state (EOS) of iron by \citet{bouchet2013ab}, the silicate-mantle model by \citet{connolly2009geodynamic} to compute equilibrium mineralogy and density profiles given the database of  \citet{stixrude2011thermodynamics}. For the water layers, we follow \citet{Vazan} using a quotidian equation of state (QEOS) and above  a pressure of 44.3 GPa, we use the tabulated EOS from \citet{seager2007}. We assume an adiabatic temperature profile within core, mantle, and water layers.

For the atmosphere, we solve the equations of hydrostatic equilibrium, mass conservation, and energy transport. For the EOS of elemental compositions of H, He, C, and O, we employ the CEA (Chemical Equilibrium with Applications) package \citep{gordon1994computer}, which performs chemical equilibrium calculations for an arbitrary gaseous mixture,  including dissociation and ionization and assuming ideal gas behavior. The envelope metallicity $Z_{\rm env}$ is the mass fraction of C and O in the envelope, which can range from 0 to 1. An irradiated atmosphere is assumed at the top of the gaseous envelope, 
for which we assume a semi-gray, analytic, global temperature averaged profile \citep{guillot2010radiative}. 
The boundary between the irradiated atmosphere and the envelope is defined where the optical depth in visible wavelength is $100 / \sqrt{3}$ \citep{jin2014planetary}. Within the envelope, the usual Schwarzschild criterion is used to distinguish between convective and radiative layers. 
The planet radius is defined where the chord optical depth becomes 0.56 \citep{Lecavelier08}. 

\begin{figure*}
\centering
 \includegraphics[width = .75\textwidth, trim = 0cm 1cm 1cm 0cm, clip]{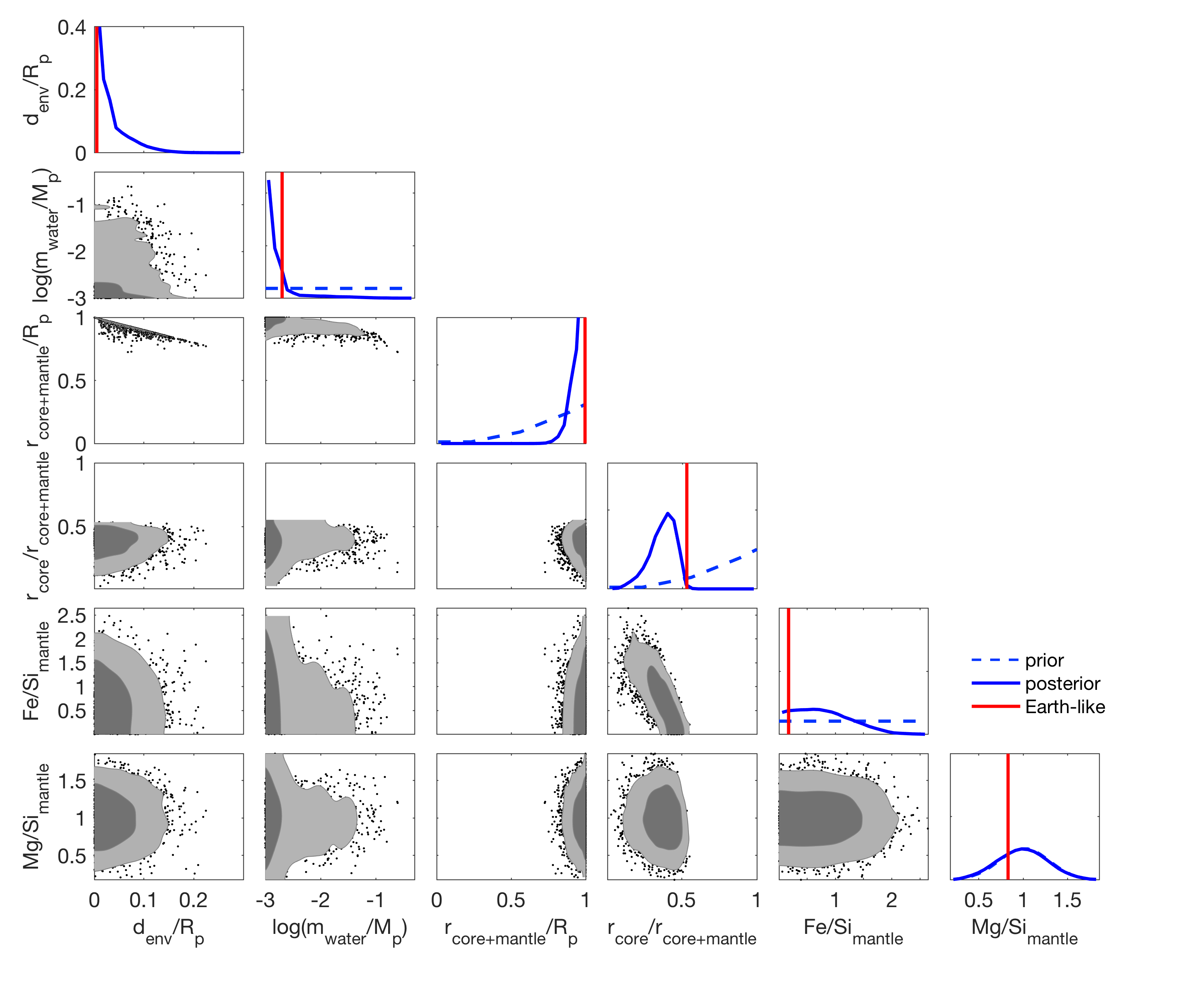}\\
 \includegraphics[width = .75\textwidth, trim = 0cm 1cm 1cm 0cm, clip]{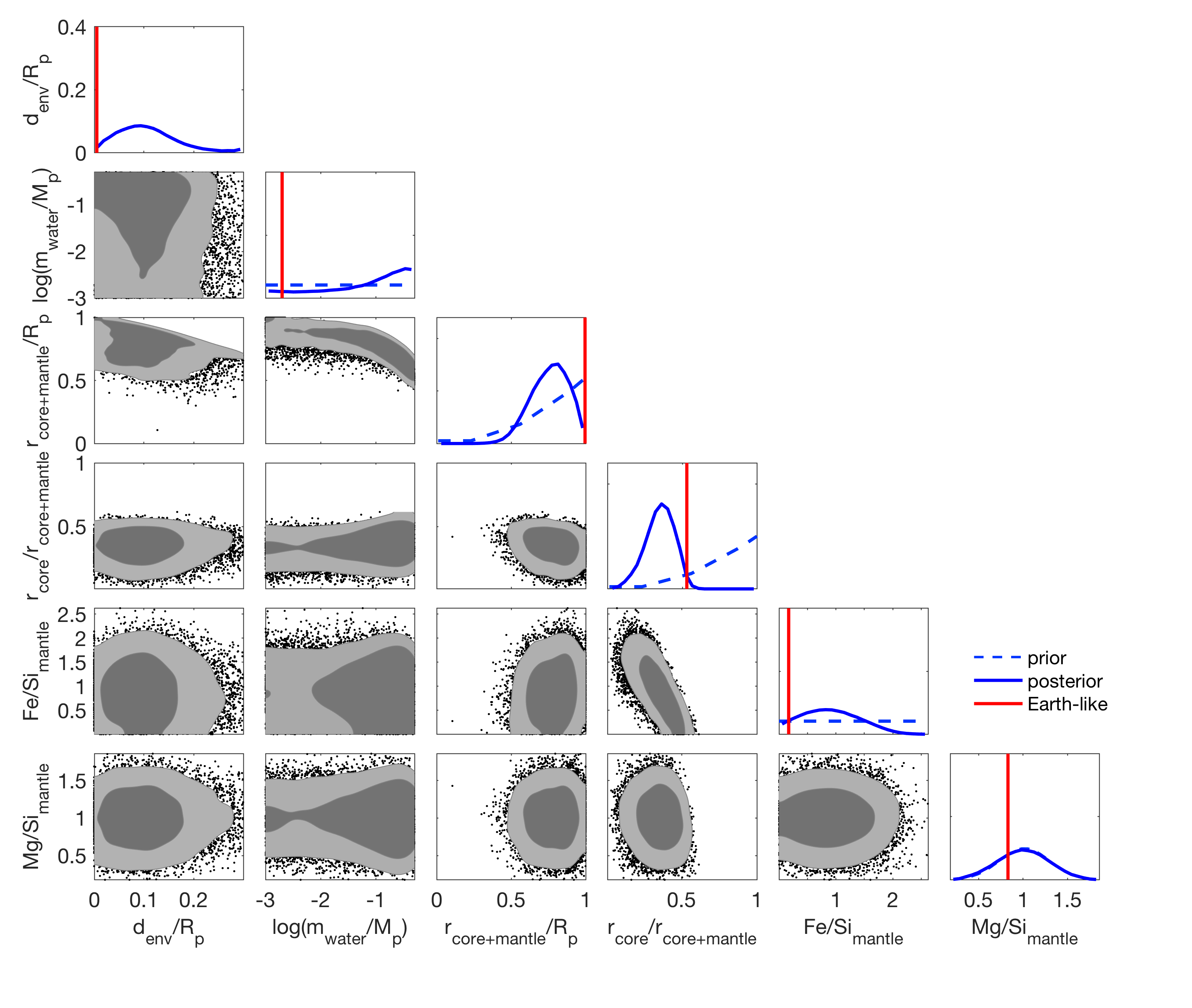}\\
 \caption{Two-and one-dimensional marginalised posterior PDFs of interior parameters for TOI-402.01 (upper panel) and TOI-402.02(lower panel): thickness of atmosphere ($d_{\rm env}$), water mass ($m_{\rm water}/{\rm M}_{\rm p}$), size of rocky interior ($r_{\rm core+mantle}$), core size ($r_{\rm core}$), and mantle composition (Fe/Si$_{\rm mantle}$ and Mg/Si$_{\rm mantle}$). The prior PDFs are shown in dashed (except for $d_{\rm env}$ for which no explicit prior is defined), while the posterior PDFs is shown in solid lines. Earth-like interior values are shown in red for reference.}
 \label{plot_interior}
\end{figure*}

\begin{table*}
\caption{Interior parameter estimates with 1$\sigma$ uncertainties of the marginalised posterior PDFs.
 \label{tableresults}}
\begin{center}
\begin{tabular}{l|lll|l}
\hline\noalign{\smallskip}
planet & \multicolumn{2}{c}{TOI-402.01} &  TOI-402.02 & \\ 
interior parameter & all constraints  &  no Fe/Si constraint &all constraints & Earth-like value\\ 
\noalign{\smallskip}
\hline\noalign{\smallskip}
log$_{10}$({m$_{\rm atm}/M_{\rm p}$})   & $-6.97_{-3.11}^{+3.35}$  & $-7.42_{-2.68}^{+3.27}$        & $-2.53_{-1.01}^{+0.87}$ &-6.06\\
\Zenv                                                           & $0.62_{-0.34}^{+0.26}$   & $0.61_{-0.33}^{+0.27}$        & $0.35_{-0.22}^{+0.32}$ &1\\
log$_{10}$(\Lenv [erg/s])                           & $21.50_{-2.10}^{+2.14}$  & $21.52_{-2.15}^{+2.19}$        & $21.61_{-2.21}^{+2.24}$  &20.6\\
$r_{\rm env}$/R$_{\rm p}$                       & $0.01_{-0.01}^{+0.05}$   & $0.01_{-0.01}^{+0.05}$        & $0.10_{-0.05}^{+0.06}$ &0.005\\
$m_{\rm water}/M_{\rm p}$                       & $0.00_{-0.00}^{+0.00}$  & $0.00_{-0.00}^{+0.00}$        & $0.10_{-0.10}^{+0.22}$ & 0.002\\
\rsolid /R$_p$                                             & $0.99_{-0.06}^{+0.01}$   & $0.99_{-0.05}^{+0.01}$        & $0.77_{-0.13}^{+0.11}$ &0.9955\\
$r_{\rm core}$/\rsolid                                 & $0.38_{-0.10}^{+0.07}$  & $0.68_{-0.16}^{+0.14}$        & $0.35_{-0.10}^{+0.09}$ &0.53\\
$\fesima$                                                      & $0.72_{-0.48}^{+0.56}$&  $7.74_{-5.16}^{+5.12}$      & $0.88_{-0.50}^{+0.54}$ &0.17\\
$\mgsima$                                                   & $1.00_{-0.27}^{+0.26}$ &  $1.00_{-0.26}^{+0.27}$      & $1.00_{-0.28}^{+0.27}$ &0.83\\
${\rho_p}/\rho_\oplus$                              & $1.25_{-0.17}^{+0.13}$    &  $1.71_{-0.31}^{+0.33}$      & $0.82_{-0.22}^{+0.25}$ & 1\\
\hline
\end{tabular} 
\end{center}
\end{table*}

\end{appendix}

\end{document}